\documentclass[twocolumn, twocolappendix]{aastex631}
\usepackage{amsmath}
\usepackage{multirow}
\usepackage{enumitem}
\usepackage{color}
\usepackage{ulem}
\usepackage{CJKfntef,CJK}
\usepackage{makecell}
\usepackage{graphicx}

\newcommand\hh{$\rm H_2$}
\newcommand\hcop{$\rm HCO^+$}

\newcommand\coa{$\rm ^{12}CO$}
\newcommand\cob{$\rm ^{13}CO$}

\newcommand\hcn{$\rm HCN$}

\newcommand\dcop{$\rm DCO^+$}
\newcommand\kms{$\rm km\ s^{-1}$}
\newcommand\hocp{$\rm HOC^+$}
\newcommand\hcsp{$\rm HCS^+$}
\newcommand\lccchp{\textit{l}-$\rm C_3H^+$}
\newcommand\cch{$\rm C_2H$}
\newcommand\cccch{$\rm C_4H$}
\newcommand\hhco{$\rm H_2CO$}
\newcommand\hcccn{$\rm HC_3N$}
\newcommand\ccthreehtwo{\textit{c-}$\rm C_3H_2$}
\newcommand\occthreehtwo{\textit{o-c-}$\rm C_3H_2$}
\newcommand\pccthreehtwo{\textit{p-c-}$\rm C_3H_2$}

\newcommand\hcoponco{$N({\rm HCO^+})/N({\rm CO})$}
\newcommand\dcoponhcop{$N({\rm DCO^+})/N({\rm HCO^+})$}
\newcommand\hcoponhocp{$N({\rm HCO^+})/N({\rm HOC^+})$}
\newcommand\hcsponcs{$N({\rm HCS^+})/N({\rm CS})$}

\newcommand\xlccchp{$X({l\text{-}\rm C_3H^+})$}

\begin{document}
\begin{CJK*}{UTF8}{gbsn}
\title{A Yebes W band Line Survey towards an Unshocked Molecular Cloud of Supernova Remnant 3C391: Evidence of Cosmic-Ray-Induced Chemistry}

\author[0000-0002-9776-5610]{Tian-yu Tu (涂天宇)}
\affiliation{School of Astronomy \& Space Science, Nanjing University, 163 Xianlin Avenue, Nanjing 210023, China}

\author[0009-0001-6483-7366]{Prathap Rayalacheruvu}
\affiliation{School of Earth and Planetary Sciences, National Institute of Science Education and Research, Jatni 752050, Odisha, India}
\affiliation{Homi Bhabha National Institute, Training School Complex, Anushaktinagar, Mumbai 400094, India}

\author[0000-0001-7031-8039]{Liton Majumdar}
\affiliation{School of Earth and Planetary Sciences, National Institute of Science Education and Research, Jatni 752050, Odisha, India}
\affiliation{Homi Bhabha National Institute, Training School Complex, Anushaktinagar, Mumbai 400094, India}

\author[0000-0002-4753-2798]{Yang Chen (陈阳)}
\affiliation{School of Astronomy \& Space Science, Nanjing University, 163 Xianlin Avenue, Nanjing 210023, China}
\affiliation{Key Laboratory of Modern Astronomy and Astrophysics, Nanjing University, Ministry of Education, Nanjing 210023, China}
\email{ygchen@nju.edu.cn}

\author[0000-0002-5683-822X]{Ping Zhou (周平)}
\affiliation{School of Astronomy \& Space Science, Nanjing University, 163 Xianlin Avenue, Nanjing 210023, China}
\affiliation{Key Laboratory of Modern Astronomy and Astrophysics, Nanjing University, Ministry of Education, Nanjing 210023, China}

\author[0000-0002-7338-0986]{Miguel Santander-Garc\'ia}
\affiliation{Observatorio Astron\'omico Nacional (OAN-IGN), Spain}

\begin{abstract}
Cosmic rays (CRs) have strong influences on the chemistry of dense molecular clouds (MCs). 
To study the detailed chemistry induced by CRs, we conducted a Yebes W band line survey towards an unshocked MC (which we named as 3C391:NML) associated with supernova remnant (SNR) 3C391. 
We detected emission lines of 18 molecular species in total and estimated their column densities with local thermodynamic equilibrium (LTE) and non-LTE analysis. 
Using the abundance ratio \hcoponco\ and an upper limit of \dcoponhcop, we estimated the CR ionization rate of 3C391:NML is $\zeta\gtrsim 2.7\times 10^{-14}\rm \ s^{-1}$ with an analytic method. 
However, we caution on adopting this value because chemical equilibrium, which is a prerequisite of using the equations, is not necessarily reached in 3C391:NML. 
We observed lower \hcoponhocp, higher \hcsponcs, and higher \xlccchp\ by an order of magnitude in 3C391:NML than the typical values in quiescent dense MCs. 
We found that an enhanced CR ionization rate (of order $\sim 10^{-15}$ or $\sim 10^{-14}\rm \ s^{-1}$) is needed to reproduce the observation with chemical model. 
This is higher than the values found in typical MCs by 2--3 orders of magnitude. 

\end{abstract}


\section{Introduction} \label{sec:intro}
Supernova remnants (SNRs) are believed to be the prime accelerator of cosmic rays (CRs) in our Galaxy \citep{Aharonian_Gamma_2013}. 
While high-energy CRs ($\gtrsim 280\rm \ MeV$) can emit $\gamma$-rays through p-p interaction with molecular clouds (MCs), low-energy CRs act as the dominating source of ionization in MCs shielded from UV radiation \citep{Spitzer_Physical_1978,Padovani_Cosmic-ray_2009}. 
CR protons can ionize molecular hydrogen through:
\begin{equation} \label{react:1}
    \rm H_2 \longrightarrow H_2^+ + e^-
\end{equation}
followed by: 
\begin{equation}
    \rm H_2^+ + H_2 \longrightarrow H_3^+ + H. 
\end{equation}
The $\rm H_3^+$ ion starts the process of the formation of polyatomic molecular species. 
The CR ionization rate per \hh, defined as the rate coefficient of reaction reaction \ref{react:1} (here we do not include the effect of secondary electrons), is an important parameter to quantify the ionization effect of CRs in a certain astrophysical environment.
Chemical effects of CRs on MCs are not limited to the ionization of $\rm H_2$. 
CRs can also drive the transition of $\rm CO\rightarrow C \rightarrow C^+$ \citep{Bisbas_Cosmic-ray_2017}. 
Other chemical effects of CRs include CR-induced non-thermal desorption \citep{Hasegawa_New_1993}, CR-induced UV photons \citep{Prasad_UV_1983a,Sternberg_Cosmic-Ray_1987}, grain sputtering \citep{Wakelam_Efficiency_2021,Paulive_modelling_2022a}, radiolysis \citep{Shingledecker_Cosmic-Ray-driven_2018,Paulive_role_2021}, etc. 

\par

Supernova remnants provide an ideal environment to study how CRs affect the chemistry in MCs. 
Observations have revealed enhanced CR ionization rate ($\sim 10^{-15}\ \rm s^{-1}$ compared with typical values $\sim 10^{-17}\rm \ s^{-1}$ \citep{Glassgold_Model_1974a}) in dense MCs associated with SNRs W51C \citep{Ceccarelli_Supernova-enhanced_2011}, W28 \citep{Vaupre_Cosmic_2014,Tu_Shock_2024}, W44 \citep{Cosentino_Interstellar_2019}, and W49B \citep{Zhou_Unusually_2022b}. 
All of these SNRs are interacting with MCs and exhibit hadronic $\gamma$-ray emission originated from the collision between the high-energy CR protons and H nucleus in the MCs. 
However, detailed behavior of CR chemistry is seldom studied in the environment of an SNR. 

\par

\begin{figure}[!t]
\centering
\includegraphics[width=0.48\textwidth]{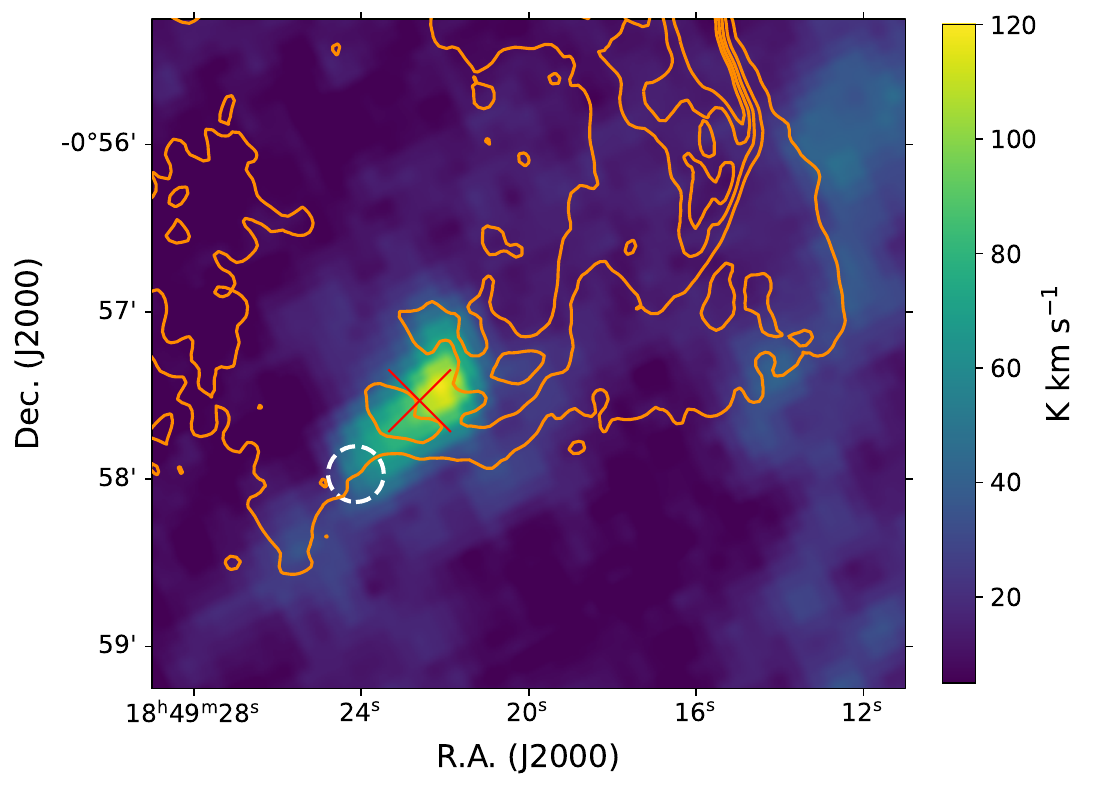}
\caption{Integrated intensity map of \coa\ $J=3$--2 line between $+100$ and $+110$ \kms\ towards the southwestern part of SNR 3C391, overlaid with orange contours of 1.4 GHz radio continuum (levels are 4, 12, 20, 28 and 36 mJy/beam). The red cross shows the 1720 MHz OH maser reported by \citet{Frail_Survey_1996}. The dashed white circle delineates the 3C391:NML region we observed. 
\label{fig:1}}
\end{figure}

SNR 3C391 is among the prototype SNRs interacting with MCs \citep{Jiang_Cavity_2010} evidenced by 1720 MHz OH masers \citep{Frail_Survey_1996}, broadened molecular lines \citep[][hereafter RR99]{Reach_Excitation_1999}, and infrared emission from molecules, atoms and ions \citep[e.g.][]{Reach_Shockingly_1996,Reach_Detection_1998,Reach_Infrared_2000a}. 
The OH masers hint enhanced CR ionization rate \citep{Nesterenok_Modelling_2022}. 
A marginal detection of the 6.4 keV Fe I K$\alpha$ line was reported by \citet{Sato_Discovery_2014}, which is likely the fluorescence produced by interaction between low-energy CRs and dense gas. 
In the $\gamma$-ray band, GeV emission has been detected with Fermi \citep{Ergin_Recombining_2014}, which is also in favor of enhanced CR ionization rate. 
In addition, an unshocked MC dominated by narrow molecular lines is found outside the shocked cloud (see Figure 6 of RR99 and Figure \ref{fig:1}), which provide an excellent site to study the chemical effect of CRs free from the disturbance of the SNR shock wave. 
Hereafter we refer to this region as 3C391:NML (narrow molecular line) following the nomenclature of RR99 who named the shocked region around the southern 1720 MHz OH maser as 3C391:BML (broad molecular line). 

\par

In this paper, we present a new W band (71.5--90 GHz) molecular line survey with the Yebes 40 m radio telescope towards 3C391:NML, aiming at enlarging the sample of SNRs exhibiting enhanced CR ionization rate and investigating the detailed CR chemistry. 
The paper is organized as follows. 
We describe the new observation and the archival data in Section \ref{sec:obs}, and present the observation results in Section \ref{sec:res}. 
In Section \ref{sec:disc}, we estimate the column densities of detected molecular species, discuss how CRs affect the observed abundances and abundance ratios, and present the result of our chemical simulation. 
The conclusions are summarized in Section \ref{sec:con}.

\section{Observations} \label{sec:obs}

\begin{figure*}[htbp]
\centering
\includegraphics[width=0.98\textwidth]{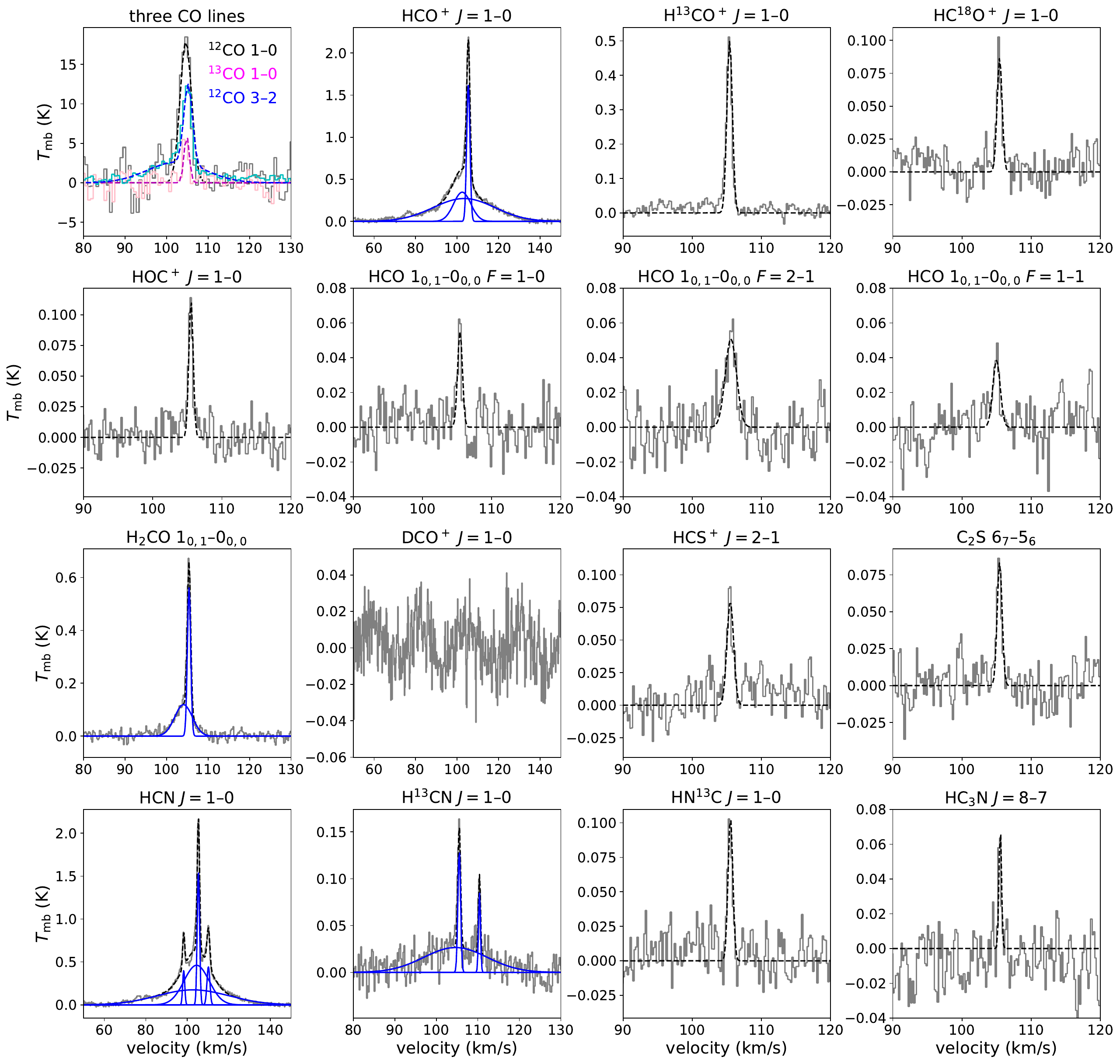}
\caption{The spectra of detected molecular lines by the Yebes 40 m observation, as well as the spectra of three CO lines and the non-detection of \dcop\ $J=1$--0 line. For the CO lines, the solid grey, pink and cyan lines show the \coa\ $J=1$--0, \cob\ $J=1$--0, and \coa\ $J=3$--2 lines, respectively, while the dashed black, magenta and blue lines show the results of (multi-)Gaussian fitting to the \coa\ $J=1$--0, \cob\ $J=1$--0, and \coa\ $J=3$--2 lines, respectively. For other molecular transitions, the grey lines shows the observed spectra, the black lines shows the fitting results, and the blue lines shows the components if multi-Gaussian fitting is adopted. The ranges of the local-standard-of-rest (LSR) velocity are adjusted for better visualization. 
\label{fig:spec1}}
\end{figure*}

\begin{figure*}[htbp]
\centering
\includegraphics[width=0.9\textwidth]{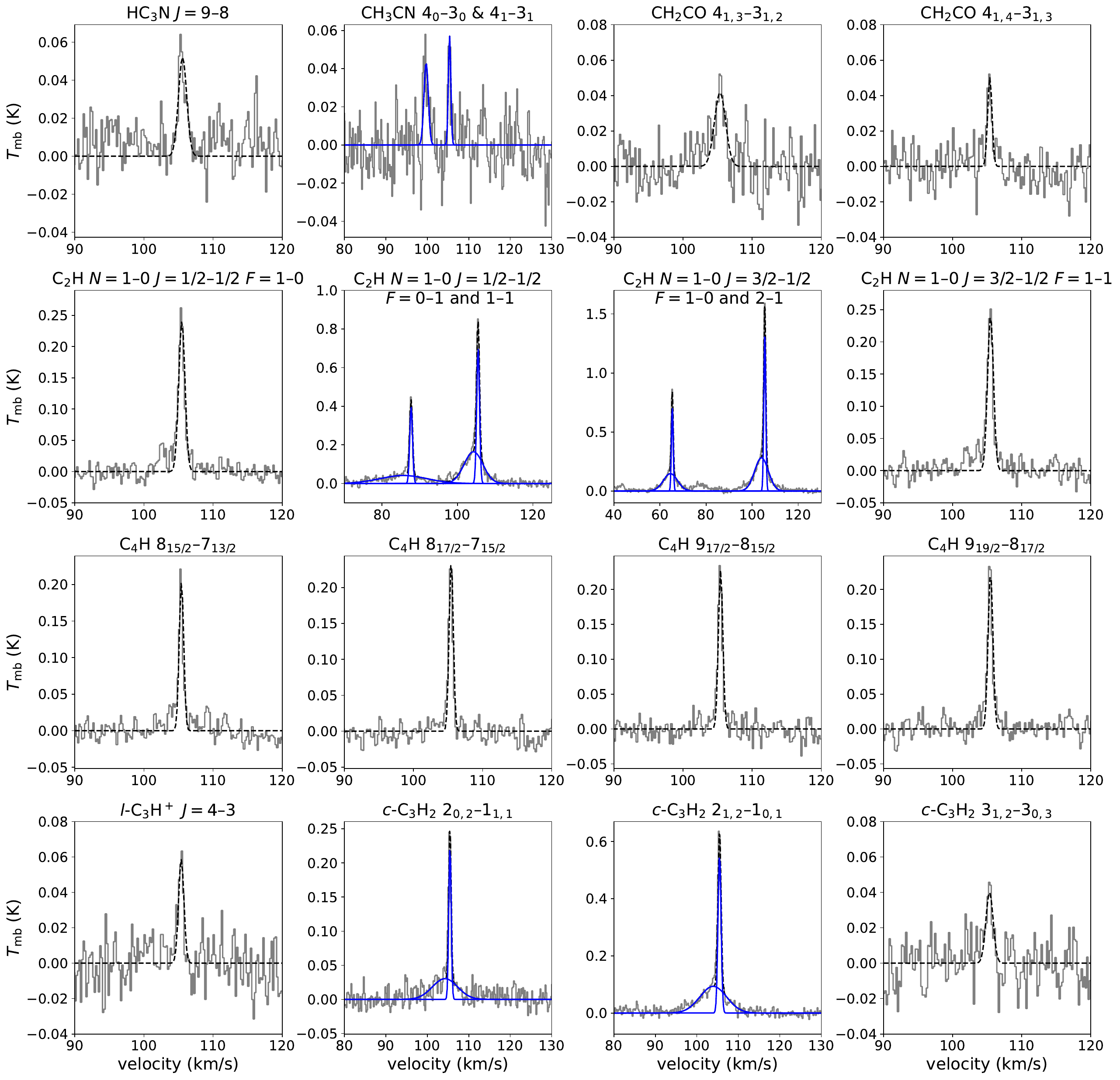}
\caption{Same as Figure \ref{fig:spec1} but for different species. 
\label{fig:spec2}}
\end{figure*}

\subsection{Yebes 40 m observation}

We carried out new pointing observation with the Yebes 40 m radio telescope (PI: Tian-Yu Tu, project code: 23A021) towards 3C391:NML ($\alpha_{J2000}={\rm 18^h49^m24^s.13}$, $\delta_{J2000}=-0^\circ57^\prime58^{\prime\prime}.17$, see the dashed white circle in Figure \ref{fig:1}), which was found by RR99 with narrow-line CS $J=2\text{--}1$ emission. 
Position switching mode was adopted throughout the observation with the reference point at $\alpha_{J2000}={\rm 18^h52^m08^s}$, $\delta_{J2000}=-1^\circ09^\prime17^{\prime\prime}$. 
The spectral coverage ranged from 71.5 to 90 GHz. 
The HPBW of the telescope was in a range of $\sim 20^{\prime\prime}$--$24^{\prime\prime}$. 
The data were smoothed to a velocity channel width of 0.2 \kms, and the resulting sensitivity measured in main beam temperature ($T_{\rm mb}$) is $\approx12$--$15\ \rm mK$ depending on the frequency. 
The raw data was reduced with the GILDAS/CLASS package\footnote{\url{https://www.iram.fr/IRAMFR/GILDAS/}}.

\subsection{Other archival data}
We used other archival data to support our analysis. 
We obtained \coa\ and \cob\ $J=1\text{--}0$ data from the FUGIN (FOREST Unbiased Galactic plane Imaging survey with the Nobeyama 45-m telescope) project \citep{Umemoto_FOREST_2017}. The angular resolutions are $20^{\prime\prime}$ for \coa\ and $21^{\prime\prime}$ for \cob, and the sensitivity estimated in $T_{\rm mb}$ is $\sim 1\text{--}3 \rm \ K$ at a velocity channel width of 0.65 \kms. 

\par

We also obtained \coa\ $J=3\text{--}2$ data from the \coa\ (3--2) High-Resolution Survey (COHRS) project \citep{Park_12CO_2023}. 
The angular resolution is $16.6^{\prime\prime}$ and the sensitivity measured in $T_{\rm A}^*$ is $\sim 1\rm \ K$ at a velocity channel width of 0.635 \kms. 
The antenna temperature was converted to $T_{\rm mb}$ with a main beam efficiency of 0.61. 
We smoothed the data to an angular resolution of $20^{\prime\prime}$ which is similar to the beam of the FUGIN data and our Yebes data. 

\par

Supplementary VLA 1.4 GHz radio continuum map was taken from the SNRcat\footnote{\url{http://snrcat.physics.umanitoba.ca}} \citep{Ferrand_census_2012}. 
All the processed data were further analyzed with \emph{Python} packages Astropy \citep{AstropyCollaboration_Astropy_2018,AstropyCollaboration_Astropy_2022} and Spectral-cube \citep{Ginsburg_Radio_2015}.
The data cubes of the CO isotopes were reprojected with Montage\footnote{\url{http://montage.ipac.caltech.edu/}} package. 
We visualized the data with \emph{Python} package Matplotlib\footnote{\url{https://matplotlib.org/}}.

\section{Results} \label{sec:res}
In Figures \ref{fig:spec1} and \ref{fig:spec2} we display the spectra of all the molecular lines detected by the Yebes 40 m observation, as well as the spectra of three CO lines and the non-detection of \dcop\ $J=1$--0 line. 
We detected 18 species (including isotopes) in total, some of which are seldom studied in the environment of SNRs. 

\par

Also shown in Figure \ref{fig:spec1} and \ref{fig:spec2} are the results (multi-)Gaussian fitting to the spectra. 
All of the detected species show narrow line emission centered at $V_{\rm LSR}\approx 105.5\ \rm km/s$, consistent with the results of RR99. 
The fitted $T_{\rm peak}$ and FWMH of the narrow components are summarized in Table \ref{tab:cden}. 

\par

We find three components in the spectrum of \hcop\ (see Figure \ref{fig:spec1}): a narrow line centered at 105.4 \kms, a moderately broadened component centered at 102.5 \kms\ with an FWHM of 9.7 \kms, and a very broad component centered at 103.4 \kms\ with an FWHM of 34.8 \kms. 
The moderately broadened component has been reported by RR99, but the broadest component is detected for the first time. 
These two broadened components may be the results of the shock of SNR propagating into different layers of 3C391:NML. 
In the outer layer, the velocity of the shock is high, resulting in a large linewidth, while the opposite when the shock wave goes deeper into the cloud. 
Detailed analysis of the shocked components is beyond the scope of this study. 

\par

Similar line profile is also found in the spectra of \hcn\ $J=1$--0 line, but two peaks can be seen in the blue and red sides of the main peak. 
These are the hyperfine structures (HFS) located at $-7.1$ \kms\ and $+4.9$ \kms\ relative to the main component. 
Spectra consisting of two components, including a narrow one and a moderately broadened one, were detected in \coa\ $J=3$--2, \hhco\ $1_{0,1}$--$0_{0,0}$, $\rm H^{13}CN$ $J=1$--0, four lines of \cch, and two lines of \ccthreehtwo. 

\par

In the following contents, we mainly focus on the narrow component because it is supposed to be free from the disturbance of the SNR shock wave. 
We note that RR99 found the $T_{\rm peak}$ of the narrow component of \hcop\ is $1.0\rm \ K$, which is smaller than the fitted value shown in Table \ref{tab:cden}. 
This is because the beam size of their observation is $27^{\prime\prime}$, which is larger than our $20^{\prime\prime}$ beam. 
Using these values, we can estimate that the angular size of the MC emitting narrow line is $\approx 27^{\prime\prime}\times \left(1.0/1.62\right)^{0.5}  \approx 21^{\prime\prime}$. 
Therefore, we assume the beam filling factor is unity in the following discussions.

\section{Discussion} \label{sec:disc}

\subsection{Estimation of molecular column density}

\subsubsection{Non-LTE estimation} \label{sec:nonLTE}

\begin{figure*}[!ht]
\centering
\includegraphics[width=0.45\textwidth]{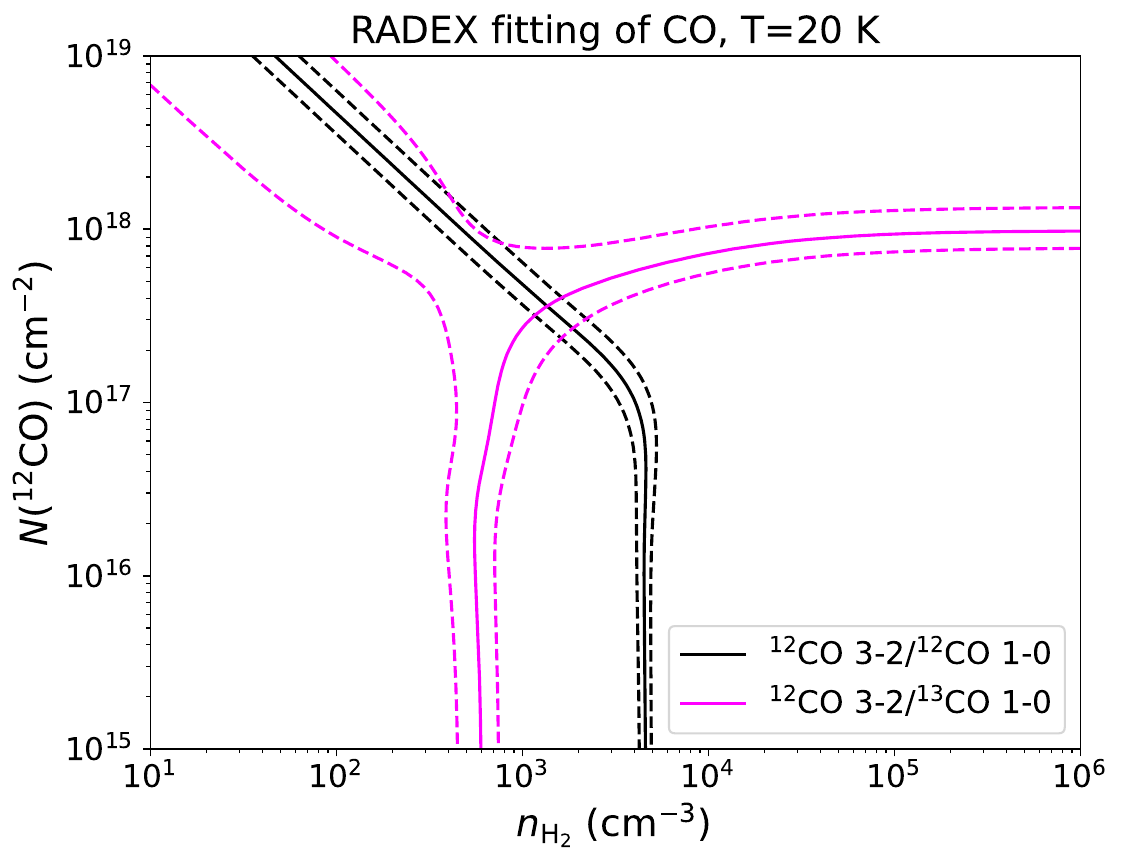}
\includegraphics[width=0.45\textwidth]{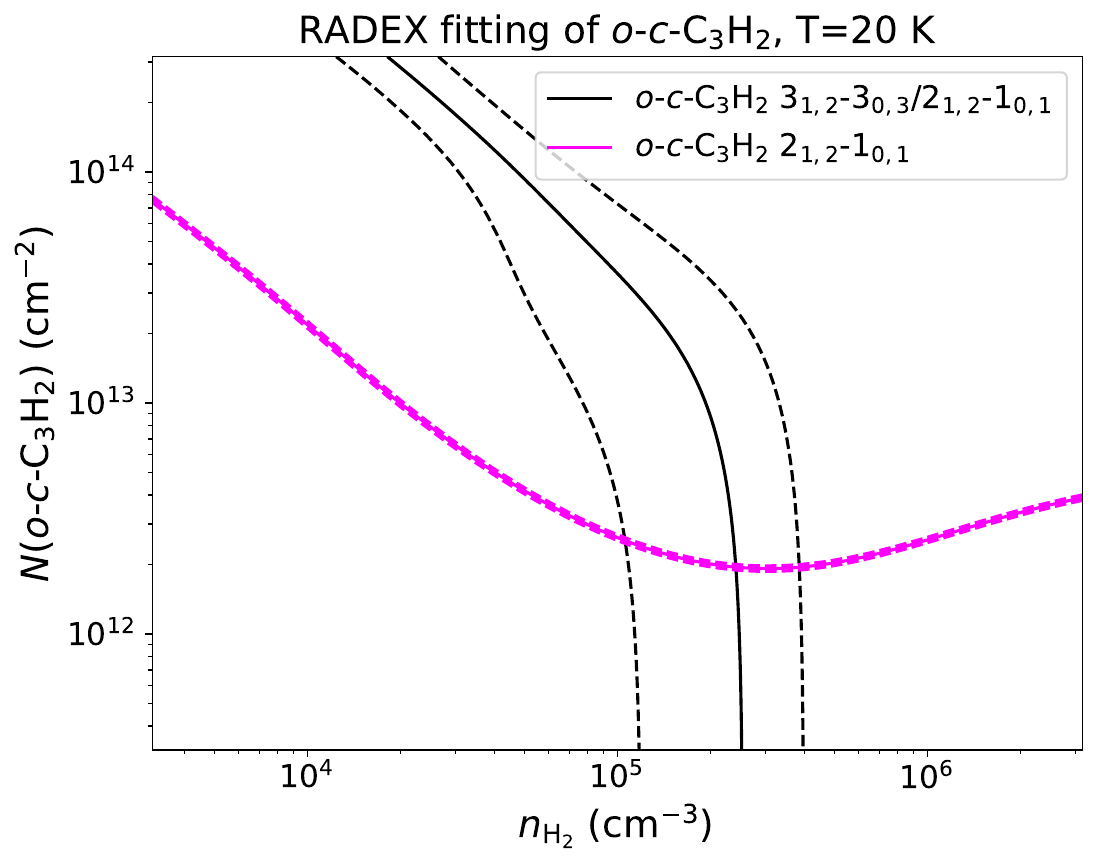}
\includegraphics[width=0.45\textwidth]{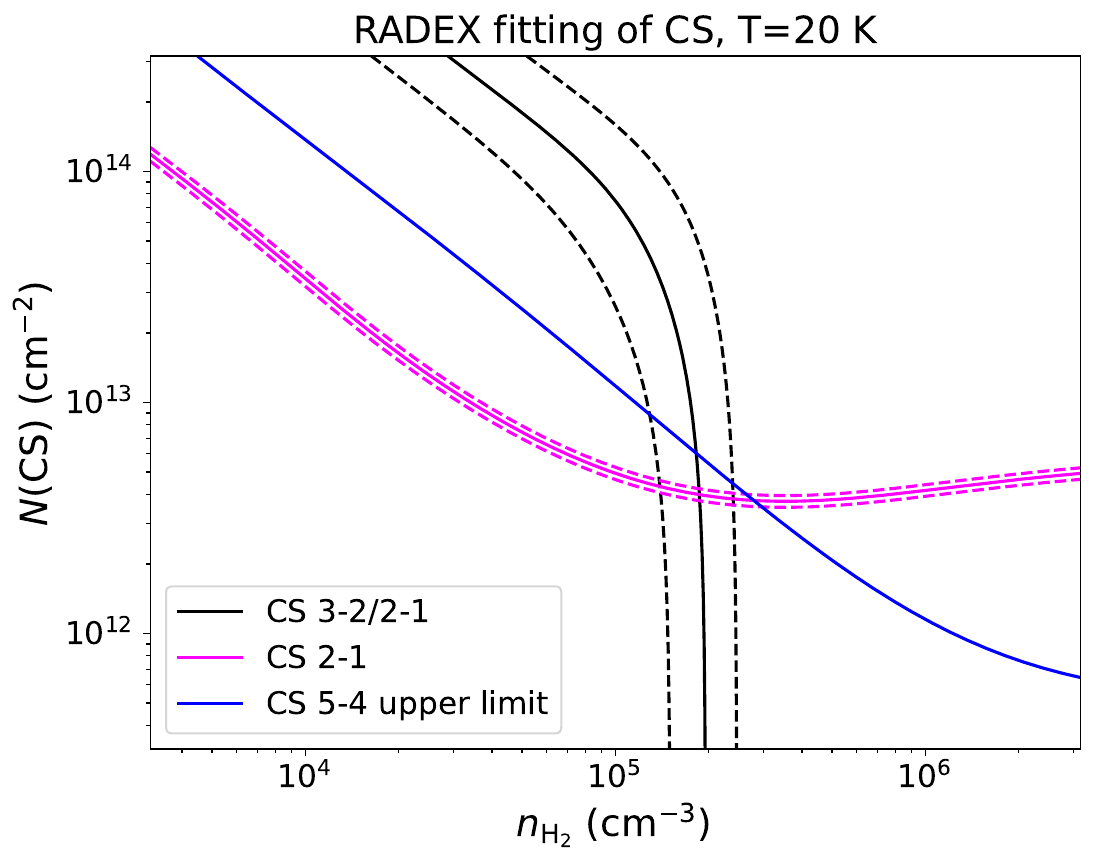}
\includegraphics[width=0.45\textwidth]{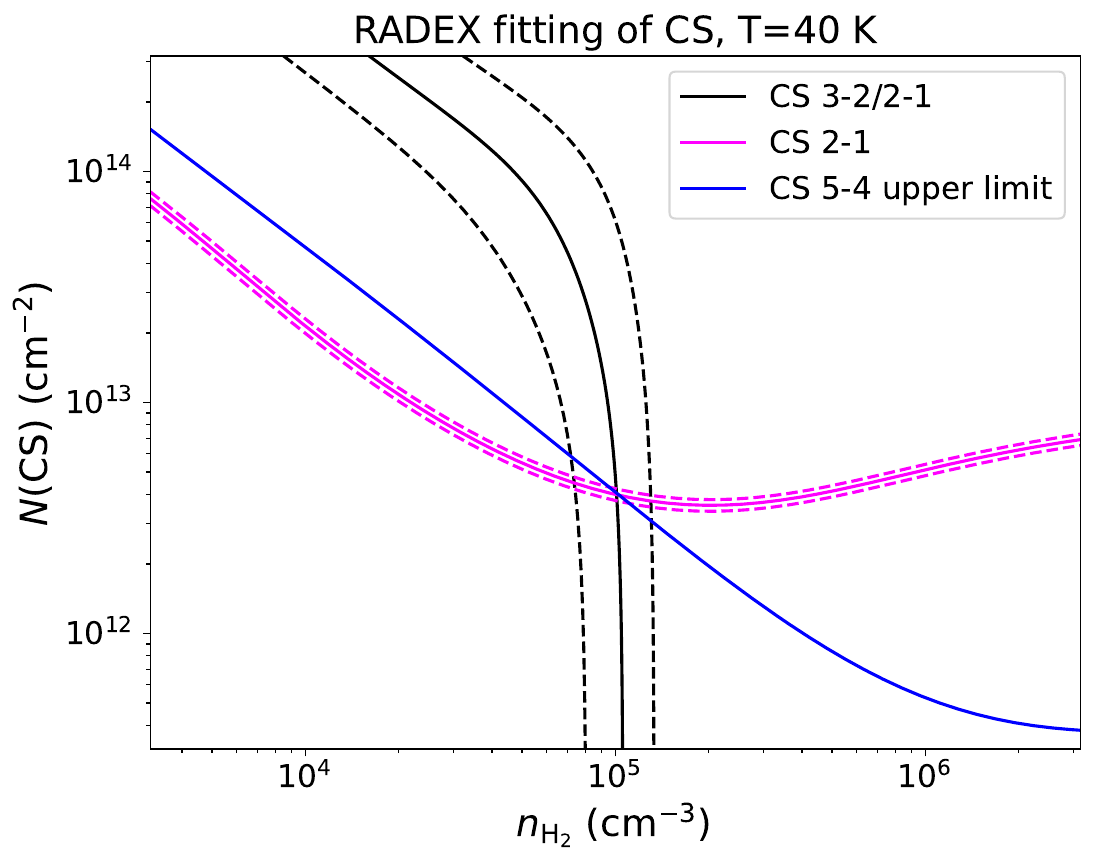}
\caption{Results of the non-LTE analysis with SpectralRadex. The solid contours show the observed $T_{\rm mb}$ or $T_{\rm mb}$ ratios of lines, while the dashed lines are the estimated uncertainties of the observed values. 
\textit{(upper left)} Results of CO with $T=20\rm \ K$. The black contours show \coa\ 3--2/\coa\ 1--0. The magenta contours show \coa\ 3--2/\cob\ 1--0. 
\textit{(upper right)} Results of \occthreehtwo\ with $T=20\rm \ K$. The black contours show \occthreehtwo\ $3_{1,2}$--$3_{0,3}$/$2_{1,2}$--$1_{0,1}$. The magenta contours show the $2_{1,2}$--$1_{0,1}$ line. 
\textit{(lower left)} Results of CS with $T=20\rm \ K$. The black contours show CS 3--2/2--1. The magenta contours show CS 2--1. The blue contour shows the upper limit of CS 5--4. 
\textit{(lower right)} Same as the lower left panel but for $T=40\rm \ K$. 
\label{fig:RADEX}}
\end{figure*}

To estimate the column densities of different molecular species of the unshocked component, we first use SpectralRadex\footnote{\url{https://spectralradex.readthedocs.io/en/latest/}}, which is a \textit{Python} wrapper of the radiative transfer code, RADEX \citep{vanderTak_computer_2007}, to present non-LTE (local thermodynamical equilibrium) analysis. 
The collisional coefficients are taken from LAMDA\footnote{\url{https://home.strw.leidenuniv.nl/~moldata/}} (Leiden Atomic and Molecular Database) \citep{vanderTak_Leiden_2020}.
The geometry of the emitting object is chosen to be sphere.
We assume a kinetic temperature $T\approx 20\rm \ K$, consistent with the value adopted by RR99 and the temperature estimated from the line ratio $I({\rm H^{13}CN})/I({\rm HN^{13}C})$ \citep{Pazukhin_H13CN-HN13C_2022}, which is $\approx 23\rm \ K$ in our case. 
The analysis is conducted with CO, \occthreehtwo, and CS. 
The data of CS emission lines are obtained from the results of RR99. 
We run a grid of density ($n_{\rm H_2}$) and column density ($N$) of specific molecular species and find the best-fit values for each species. 
The results are shown in Figure \ref{fig:RADEX} and Table \ref{tab:nonLTE}. 

\par

For CO, we use the \coa\ 1--0, \cob\ 1--0 and \coa\ 3--2 lines for the analysis, and we fit the ratios $T_{\rm mb}$(\coa\ 3--2)/$T_{\rm mb}$(\coa\ 1--0) and $T_{\rm mb}$(\coa\ 3--2)/$T_{\rm mb}$(\cob\ 1--0). 
We assume $N(^{12}{\rm CO})/N(^{13}{\rm CO})=45$ in our calculation \citep{Milam_12C_2005}. 
The best-fit values, marked by the intersection point of the two solid lines in the upper left panel of Figure \ref{fig:RADEX}, are $N({\rm ^{12}CO})\sim 3.7 \times 10^{17}\rm \ cm^{-2}$ and $n_{\rm H_2}\sim 1.4\times 10^3\rm \ cm^{-3}$. 
We note that there is parameter degeneracy in our fitting, so the values are just a rough estimation. 

\par

For \occthreehtwo, we use the \occthreehtwo\ $3_{1,2}$--$3_{0,3}$ and $2_{1,2}$--$1_{0,1}$ lines, and fit the $T_{\rm mb}(3_{1,2}$--$3_{0,3})$/$T_{\rm mb}(2_{1,2}$--$1_{0,1})$ ratio and $T_{\rm mb}(2_{1,2}$--$1_{0,1})$. The best-fit values are $N($\occthreehtwo$)\sim2.0\times 10^{12} \rm \ cm^{-2}$ and $n_{\rm H_2}\sim 2.4\times 10^5\rm \ cm^{-3}$

\par

For CS, we fit $T_{\rm mb}$(CS 3--2)/$T_{\rm mb}$(CS 2--1) and $T_{\rm mb}$(CS 2--1), and use the non-detection of CS 5--4 reported by RR99 as an upper limit. 
The best-fit values are $N({\rm CS})\sim 3.9 \times 10^{12}\rm \ cm^{-2}$ and $n_{\rm H_2}\sim 1.9\times 10^5\rm \ cm^{-3}$. 
We also show the fitting results of CS assuming $T=40\rm \ K$ in the lower right panel of Figure \ref{fig:RADEX}. 
The results shows that the kinetic temperature of 3C391:NML should not be much higher than 40 K, otherwise the CS 5--4 line should have been detected. 
The CS lines set an upper limit for the kinetic temperature $T$. 
We also note that although the kinetic temperatures are different, the best-fit $N(\rm CS)$ does not change significantly, and the variation of the corresponding excitation temperature of CS 2--1 line ($\approx 10\rm \ K$) is also not prominent. 
This excitation temperature will be used to conduct LTE analysis for the other species in the following section. 

\par

The densities estimated with the \occthreehtwo\ and CS lines are similar and higher respectively than that estimated with CO.
This is not surprising because the critical densities are $\sim 10^{6}\rm \ cm^{-3}$ for the \occthreehtwo\ and CS lines, while $\sim 10^3 $--$10^4 \rm cm^{-3}$ for the CO lines, which means that \occthreehtwo\ and CS trace a denser part of the MC than CO. 

\par

For other molecular species, non-LTE analysis is not available. 
We use LTE analysis to estimate their column densities. 

\begin{deluxetable}{ccc}[!t]
\tablecaption{Results of non-LTE analysis of CO, \occthreehtwo, and CS lines. 
\label{tab:nonLTE}}
\tablehead{
\colhead{Species} & \colhead{Column density ($\rm cm^{-2}$)} & \colhead{$n_{\rm H2}\rm \ (cm^{-3})$}
}
\startdata
CO & $\sim 3.7\times 10^{17}$ & $\sim 1.4\times 10^3$ \\
\occthreehtwo & $\sim 2.0\times 10^{12}$ & $\sim 2.4\times 10^5$ \\
CS (20 K) &$\sim 3.9\times 10^{12}$ & $\sim 1.9\times 10^5$ \\
CS (40 K) &$\sim 3.9\times 10^{12}$ & $\sim 1.0\times 10^5$
\enddata
\end{deluxetable}

\subsubsection{LTE estimation}
We assume LTE condition for the molecular species other than CO, \occthreehtwo, and CS analyzed above
Assuming the lines to be optical thin, we obtain the column densities through (Equation C1 in \citet{Liu_Search_2021}): 
\begin{equation} \label{eq:N}
\begin{split}
    N = \frac{3k}{8\pi^3\nu} \frac{Q_{\rm rot} \exp{(E_{\rm u}/kT_{\rm ex})}}{S\mu^2}& \\ 
    \times \frac{J_\nu(T_{\rm ex})}{J_\nu(T_{\rm ex})-J_\nu(T_{\rm bg})} &\int T_{\rm mb}\, dv,
\end{split}
\end{equation}
where $Q_{\rm rot}$ is the partition function, $E_{\rm u}$ is the energy of the upper level, $T_{\rm ex}$ is the excitation temperature, $S$ is the line strength, $\mu$ is the dipole moment, $T_{\rm bg}=2.73\ \rm K$ is the background temperature, and $J_\nu(T)$ is defined by $J_\nu(T)=(h\nu/k)/(\exp{(h\nu/kT)}-1)$. 
We adopt the $Q_{\rm rot}$, $E_{\rm u}$ and $S\mu^2$ from Splatalogue\footnote{\url{https://splatalogue.online/}}. 
The data of HCO are extracted from the JPL database\footnote{\url{https://spec.jpl.nasa.gov/}}, while other data are from the CDMS database\footnote{\url{https://cdms.astro.uni-koeln.de/}}. 

\par 

We use two values of $T_{\rm ex}$ (5 K and 10 K) to estimate the column density for each molecular species, which are tabulated in Table \ref{tab:cden}. 
$T_{\rm ex}$ is calculated with the following two methods. 
(1) The optical depth of \hcop\ can be obtained by solving:
\begin{equation} 
    \frac{1-{\rm e}^{-\tau}}{1-{\rm e}^{-\tau/327}} = \frac{W({\rm HCO^+})}{W({\rm HC^{18}O^+})},
\end{equation}
where $W$ is the integrated intensity and we assume $N({\rm HCO^+})/N({\rm HC^{18}O^+}) \approx  N{\rm ^{16}O}/N{\rm ^{18}O} \approx 327$ \citep{Yan_Direct_2023} considering the galactocentric distance of 3C391 is $\approx 4.4$ kpc \citep{Ranasinghe_Distances_2022b}. 
We get $\tau \approx 8$, which is high enough for us to estimate the excitation temperature of \hcop\ by \citep{Mangum_How_2015}: 
\begin{equation}
    T_{\rm ex} = \frac{h\nu/k}{\ln{(1+\frac{h\nu/k}{T_{\rm peak}/f+J_\nu(T_{\rm bg})})}}.
\end{equation}
Then we get $T_{\rm ex}=5\rm \ K$ for \hcop. 
We adopt this value as a first estimation of $T_{\rm ex}$.
(2) According to the results of RADEX (in Section \ref{sec:nonLTE}), the excitation temperature of the CS 2--1 line is $\approx 10\rm \ K$. 
We adopt this value as the second estimation of $T_{\rm ex}$. 

\par

Equation \ref{eq:N} only applies when the transition is optical thin which is, however, not necessarily satisfied in some of the detected lines. 
To minimize the influence of finite optical depth, we use the emission from rarer isotopes or the HFS to calculate the column density when isotopes or HFS are detected. 
Specifically, we use $\rm HC^{18}O^+$ to calculate $N(\rm HCO^+)$, $\rm H^{13}CN$ for $N(\rm HCN)$, and \cch\ $N=1$--0 $J,F=3/2,1$--1/2,1 for $N({\rm C_2H})$. 
For \cccch, the results estimated with the two 8--7 lines (or the two 9--8 lines) are similar, but there is a significant difference between the values estimated by 8--7 and 9--8 lines. 
For HCO, \hcccn, $\rm CH_2CO$ and $\rm CH_3CN$, although multiple transitions are detected, we calculate the column densities with all detected transitions because they are rather weak and the results are different. 

\par

As seen in Table \ref{tab:cden}, for most species, the column densities estimated with the two different $T_{\rm ex}$ are consistent within a factor of 2, except \hcccn, \cccch, $\rm CH_2CO$ and $\rm CH_3CN$. 
For \hcccn, $\rm CH_2CO$ and $\rm CH_3CN$, there is also a non-negligible difference between the estimated column densities with different transitions, probably because the lines are rather weak and the relative uncertainty is high.

\begin{deluxetable*}{ccccccc}[ht]
\tablecaption{Results from Gaussian fitting of the narrow line components of all detected lines in the Yebes 40 m observation, as well as the column densities of molecular species. 
\label{tab:cden}}
\tablehead{ 
\colhead{Species} & \colhead{Transition} & 
\colhead{\makecell{Frequency \\ (MHz)}} & \colhead{\makecell{$T_{\rm peak}$ \\ (K)}} & 
\colhead{\makecell{FWHM \\ ($\rm km\ s^{-1}$)}} & 
\colhead{\makecell{$N(T_{\rm ex}=5\rm \ K)$$\rm ^a$ \\ $\rm (cm^{-2})$}} & 
\colhead{\makecell{$N(T_{\rm ex}=10\rm \ K)$$\rm ^a$ \\ $\rm (cm^{-2})$}}}
\startdata 
$\rm HCO^+$ & 1--0 & 89188.525 & 1.62 & 1.87 & \multirow{3}{*}{\begin{tabular}[c]{@{}c@{}}3.3(13)\\ for $\rm HCO^+$$\rm ^b$\end{tabular}} & \multirow{3}{*}{\begin{tabular}[c]{@{}c@{}}2.9(13)\\ for $\rm HCO^+$$\rm ^b$\end{tabular}} \\
$\rm H^{13}CO^+$ & 1--0 & 86754.288 & 0.50 & 0.93 &  &  \\
$\rm HC^{18}O^+$ & 1--0 & 85162.223 & 0.087 & 0.74 &  &  \\ \hline
$\rm DCO^+$ & 1--0 & 72039.312 & $<0.045$$^c$ & --- & $<7.7$(10) & $<7.1$(10) \\ \hline
\multirow{3}{*}{$\rm HCN$} & 1--0 $F=0$--1 & 88633.936 & 0.40 & 1.16 & \multirow{5}{*}{\begin{tabular}[c]{@{}c@{}}1.5(13)\\ for HCN$\rm ^d$\end{tabular}} & \multirow{5}{*}{\begin{tabular}[c]{@{}c@{}}1.3(13)\\ for HCN$\rm ^d$\end{tabular}} \\
 & 1--0 $F=2$--1 & 88631.847 & 1.54 & 1.36 &  &  \\
 & 1--0 $F=1$--1 & 88630.416 & 0.44 & 1.63 &  &  \\ \cline{1-5}
\multirow{2}{*}{$\rm H^{13}CN$} & 1--0 $F=1$--1 & 86338.737 & 0.085 & 0.56 &  &  \\
 & 1--0 $F=2$--1 & 86340.176 & 0.13 & 0.77 &  &  \\ \hline
$\rm HN^{13}C$ & 1--0 & 87090.859 & 0.10 & 0.81 & 1.8(11) & 1.6(11) \\ \hline
$\rm HOC^+$ & 1--0 & 89487.414 & 0.11 & 0.73 & 2.1(11) & 1.9(11) \\ \hline
$\rm HCO$ & $1_{0,1}$--$0_{0,0}$ & \multicolumn{1}{l}{} & \multicolumn{1}{l}{} & \multicolumn{1}{l}{} & \multicolumn{1}{l}{} & \multicolumn{1}{l}{} \\
 & $J,F=3/2,1$--1/2,0 & 86708.35 & 0.056 & 0.80 & 1.2(12) & 2.0(12) \\
 & $J,F=3/2,2$--1/2,1 & 86670.82 & 0.050 & 1.96 & 1.6(12) & 2.5(12) \\
 & $J,F=1/2,1$--1/2,1 & 86777.43 & 0.038 & 1.28 & 1.4(12) & 2.2(12) \\ \hline
$\rm H_2CO$ & $1_{0,1}$--$0_{0,0}$ & 72837.951 & 0.57 & 0.93 & 3.5(12) & 6.3(12) \\ \hline
$\rm HCS^+$ & 2--1 & 85347.869 & 0.080 & 1.13 & 6.9(11) & 5.3(11) \\ \hline
\multirow{2}{*}{$\rm HC_3N$} & 8--7 & 72783.818 & 0.075 & 0.41 & 5.9(11) & 1.7(11) \\
 & 9--8 & 81881.463 & 0.052 & 1.40 & 2.4(12) & 5.5(11) \\ \hline
$\rm C_2H$ & $N=1$--0 & \multicolumn{1}{l}{} & \multicolumn{1}{l}{} & \multicolumn{1}{l}{} & \multicolumn{1}{l}{} & \multicolumn{1}{l}{} \\
 & $J,F=1/2,1$--1/2,0 & 87446.512 & 0.24 & 1.08 & \multirow{6}{*}{1.9(14)} & \multirow{6}{*}{1.8(14)} \\
 & $J,F=1/2,0$--1/2,1 & 87407.165 & 0.39 & 0.88 &  &  \\
 & $J,F=1/2,1$--1/2,1 & 87402.004 & 0.70 & 0.89 &  &  \\
 & $J,F=3/2,1$--1/2,0 & 87328.624 & 0.70 & 0.92 &  &  \\
 & $J,F=3/2,2$--1/2,1 & 87316.925 & 1.34 & 0.99 &  &  \\
 & $J,F=3/2,1$--1/2,1 & 87284.156 & 0.24 & 1.01 &  &  \\ \hline
$\rm C_2S$ & $6_7$--$5_6$ & 81505.208 & 0.086 & 0.80 & 1.6(12) & 8.8(11) \\ \hline
\multirow{4}{*}{$\rm C_4H$} & $8_{17/2}$--$7_{15/2}$ & 76117.43 & 0.23 & 0.73 & \multirow{2}{*}{4.3(13)} & \multirow{2}{*}{1.2(13)} \\
 & $8_{15/2}$--$7_{13/2}$ & 76156.02 & 0.20 & 0.77 &  &  \\ \cline{2-7} 
 & $9_{19/2}$--$8_{17/2}$ & 85634.00 & 0.22 & 0.81 & \multirow{2}{*}{8.3(13)} & \multirow{2}{*}{1.6(13)} \\
 & $9_{17/2}$--$8_{15/2}$ & 85672.57 & 0.23 & 0.76 &  &  \\ \hline
\lccchp & 4--3 & 89957.625 & 0.067 & 0.75 & 3.5(11) & 1.8(11) \\ \hline
\multirow{2}{*}{\occthreehtwo} & $3_{1,2}$--$3_{0,3}$ & 82966.201 & 0.041 & 1.22 & \multicolumn{2}{c}{\multirow{2}{*}{2.0(12)$\rm ^e$}} \\
 & $2_{1,2}$--$1_{0,1}$ & 85338.906 & 0.56 & 0.87 & \multicolumn{2}{c}{} \\ \hline
\pccthreehtwo & $2_{0,2}$--$1_{1,1}$ & 82093.555 & 0.23 & 0.68 & 3.7(12) & 4.1(12) \\ \hline
\multirow{2}{*}{$\rm CH_2CO$} & $4_{1,3}$--$3_{1,2}$ & 81586.229 & 0.041 & 1.92 & 1.7(13) & 4.8(12) \\
 & $4_{1,4}$--$3_{1,3}$ & 80076.644 & 0.051 & 0.74 & 8.1(12) & 2.3(12) \\ \hline
\multirow{2}{*}{$\rm CH_3CN$} & $4_1$--$3_1$ & 73588.799 & 0.056 & 0.71 & 9.4(11) & 3.4(11) \\
 & $4_0$--$3_0$ & 73590.217 & 0.044 & 1.19 & 2.8(11) & 2.1(11) \\ \hline
\enddata
\tablecomments{
$^{\rm a}$ m(n) means $\rm m\times 10^{n}$.
$^{\rm b}$ The column density of \hcop\ is estimated from $\rm HC^{18}O^+$ assuming $N({\rm HCO^+})/N({\rm HC^{18}O^+}) \approx \rm {^{16}O}/{^{18}O} \approx 327$ \citep{Yan_Direct_2023}. 
$^{\rm c}$ \dcop\ is not detected, so we use the $3\sigma$ value as an upper limit. 
$^{\rm d}$ The column density of \hcn\ is estimated from $\rm H^{13}CN$ assuming $N({\rm HCN})/N({\rm H^{13}CN}) \approx \rm {^{12}C}/{^{13}C} \approx 45$ \citep{Milam_12C_2005}. 
$^{\rm e}$ The column density of \occthreehtwo\ is estimated with non-LTE method (see Section \ref{sec:nonLTE}). 
}
\end{deluxetable*}

\subsection{Relation between the observed molecular abundances and CR chemistry}

In Section \ref{sec:nonLTE}, we obtained the $N(^{12}{\rm CO})\sim 3.7\times 10^{17} \rm \ cm^{-2}$. 
Assuming $N({\rm H_2})\approx 7\times 10^5 N(^{13}{\rm CO})$ \citep{Frerking_relationship_1982}, we obtain $N({\rm H_2})\sim 5.8\times 10^{21}\rm \ cm^{-2}$. 
We will use this value to estimate the molecular abundance relative to \hh. 

\subsubsection{Estimation of CR ionization rate with analytic method}

The abundance ratios $R_{\rm D}=N({\rm DCO^+})/N({\rm HCO^+})$ and $R_{\rm H}=N({\rm HCO^+})/N({\rm CO})$ has been used to estimate the CR ionization rate in MCs \citep{Caselli_Ionization_1998,Ceccarelli_Supernova-enhanced_2011,Vaupre_Cosmic_2014}. 
Assuming the chemistry of 3C391:NML has reached equilibrium and the temperature is low ($\ll 220\ \rm K$), the CR ionization $\zeta$ can be calculated simply with \citep{Vaupre_Cosmic_2014}: 
\begin{equation} \label{eq:CRIR}
    \begin{split}
        \frac{\zeta}{n_{\rm H}} & = \frac{\beta^\prime}{k_{\rm H}}\left( 2\beta x_{\rm e}+\delta \right)R_{\rm H}x_{\rm e}, \\
        x_{\rm e} & = \left( \frac{k_{\rm f}x({\rm HD})}{3R_{\rm D}} - \delta \right)\frac{1}{k_{\rm e}}
    \end{split}
\end{equation}
where $x_{\rm e}$ is the ionization fraction in the MC, $x(\rm HD)$ is the abundance of HD relative to H, $k_{\rm f}$, $\delta$, $\beta$ and $\beta^{\prime}$ are rate coefficients of chemical reactions listed in the Table A.1. of \citet{Vaupre_Cosmic_2014}. 
These equations allow us to set a lower limit for $\zeta$ even though \dcop\ is not detected. 
Assuming $x(\rm HD)=1.6\times 10^{-5}$ \citep{Linsky_What_2006} and adopting $n_{\rm H}=2.8\times 10^3 \rm \ cm^{-3}$ from the non-LTE analysis of CO, we get $\zeta \gtrsim 2.7\times 10^{-14}\rm \ s^{-1}$. 
This value is higher than the typical value in MCs \citep[$\sim 10^{-17} \rm \ s^{-1}$][]{Glassgold_Model_1974a} by three orders of magnitude, and the values obtained in W51C and W28 with the same method by one order of magnitude \citep{Ceccarelli_Supernova-enhanced_2011,Vaupre_Cosmic_2014}. 

\par

We note, however, that chemical equilibrium is not necessarily reached in 3C391:NML. 
Although the timescale of gas-phase ion-neutral reactions induced by CRs is only $\sim 10^2 \ \rm yr$ \citep{Vaupre_Cosmic_2014}, grain chemistry may take longer time to reach equilibrium. 
The deuterium chemistry is strongly affected by grain processes \citep[e.g.,][]{Feng_Chemical_2020}. 
We recall that the rate coefficient of CR-induced non-thermal desorption for a specific grain species is \citep{Hasegawa_New_1993}:
\begin{equation} \label{eq:CRdes}
    k_{\rm CR} = f({\rm 70 \ K})k_{\rm des}({\rm 70 \ K}),
\end{equation}
where $f({\rm 70 \ K})$ is the fraction of the time spent by dust grains at the dust temperature $T_{\rm dust}=70 \rm \ K$ due to CR heating, which is approximately \citep{Reboussin_Grain-surface_2014}:
\begin{equation} \label{eq:CRfrac}
    f({\rm 70 \ K}) = \frac{\zeta}{1.3\times 10^{-17}\ \rm s^{-1}} 3.16\times 10^{-19}. 
\end{equation}
The value $k_{\rm des}({\rm 70 \ K})$ is the rate coefficient of thermal desorption rate at $T_{\rm dust}=70\rm \ K$, which can be expressed as: 
\begin{equation} \label{eq:thder}
    k_{\rm des} = \nu_0\exp{\left( -\frac{E_{\rm D}}{kT_{\rm dust}} \right)},
\end{equation}
where $E_{\rm D}$ is the desorption energy of the adsorbed species, and 
\begin{equation} \label{eq:nu0}
    \nu_0 = \left( \frac{2n_{\rm s}E_{\rm D}}{\pi^2m} \right) ^{1/2},
\end{equation}
where $n_{\rm s}\sim 1.5\times 10^{15}\rm \ cm^{-2}$ is the surface density of sites and $m$ is the mass of the adsorbed species. 
For CO, the desorption energy is $\approx 1200 \rm \ K$ \citep{Hasegawa_New_1993}. 
Substituting the values into these equations, we estimate that the CR-induced non-thermal desorption timescale for CO ice at $\zeta\sim 10^{-14}\rm \ s^{-1}$ is $\tau_{\rm CR}\approx 1/k_{\rm CR} \sim 3\times 10^3\rm \ yr$, which is comparable to the age of SNR 3C391 (4--19 kyr. See our further discussion in Section \ref{sec:model}). 
We note that $k_{\rm CR}$ depends exponentially on the desorption energy $E_{\rm D}$ (see Equation \ref{eq:thder}) which varies with species. 
A substantial proportion of ice is composed of $\rm H_2O$, $\rm NH_3$, $\rm CH_3OH$ etc. \citep{Ruaud_Gas_2016a}. 
These species have higher values of $E_{\rm D}$, which result in longer desorption timescale. 
Therefore, the chemistry may be highly dynamic in the environment of 3C391, and the CR ionization rate estimated with Equations \ref{eq:CRIR} may thus deviate from the real value. 

\subsubsection{Unusual abundance and abundance ratios found by the Yebes observation} \label{sec:abun}

Using the column densities listed in Table \ref{tab:cden} and the estimated $N({\rm H_2})\sim 5.8\times 10^{21}\rm \ cm^{-2}$, we can obtain the abundances of the detected species. 
We find that the obtained \hcoponhocp, \hcsponcs\ and \xlccchp\ are different from typical values found in dense MCs. 
Here $X$ denotes the abundance of a specific species relative to \hh. 
All of the three values can probably be attributed to the chemistry of CRs. 

\par

The observed value of \hcoponhocp\ is $\sim 160$--180, while typical value found in quiescent dense MCs in our Galaxy is $\sim 10^3$ \citep{Apponi_New_1997}, which is higher than our observed value by an order of magnitude. 
The low \hcoponhocp\ has been found in Galactic photodissociation regions (PDRs) (e.g. the Horsehead PDR \citep{Goicoechea_ionization_2009a} and Orion Bar PDR \citep{Goicoechea_Spatially_2017}), diffuse clouds \citep{Liszt_abundance_2004}, and extragalactic sources (e.g. NGC 253 \citep{Harada_Starburst_2021} and M 82 \citep{Fuente_chemistry_2008}). 
In low-temperature molecular gas, \citet{Harada_Starburst_2021} proposed that the decrease \hcoponhocp\ at high visual extinction $A_{\rm V}$ gas shielded from UV radiation is caused by the extremely high CR ionization rate $\sim 10^{-14}\rm \ s^{-1}$ in the central molecular zone (CMZ) of NGC 253. 
In this case, the reduction of \hcoponhocp\ is due to the enrichment of $\rm C^+$ and $\rm CO^+$, which in turn leads to a faster production of \hocp\ through reactions:
\begin{equation}\label{react:c++h2o}
  \begin{split}
    \rm C^+ + H_2O & \rm \longrightarrow HOC^+ + H \\
    & \rm \longrightarrow HCO^+ + H
  \end{split}
\end{equation}
with a branching ratio of 2:1 \citep{Jarrold_Reanalysis_1986}, and 
\begin{equation}\label{react:co++h2}
  \begin{split}
    \rm CO^+ + H_2 & \rm \longrightarrow HOC^+ + H \\
    & \rm \longrightarrow HCO^+ + H 
  \end{split}
\end{equation}
with a branching ratio of approximately 1:1 \citep{Anicich_Evaluated_1993}. 

\par

The observed value of \hcsponcs\ is $\sim 0.14$--0.18, while the typical values found in the Taurus, Perseus, Orion MCs and Barnard 1 dark cloud are all $\sim 10^{-2}$ \citep{Rodriguez-Baras_Gas_2021a,Fuente_Ionization_2016a}, which is lower than our observed value by an order of magnitude. 
The chemical simulation of \citet{Fuente_Ionization_2016a} shows that the enhanced \hcsponcs\ ratio could be a tracer of high CR ionization rate at chemical equilibrium, because $\rm H_3^+$, \hcop, and $\rm H_3O^+$, which are all important products of CR ionization, are important reactants that transform CS to \hcsp\ \citep{Podio_Molecular_2014a}. 
Though, this simulation considered only a narrow range of CR ionization rate and is limited to chemical equilibrium. 

\par

The observed value of \xlccchp\ is $\sim 3.1$--$6.0\times 10^{-11}$. 
The \lccchp\ molecular species is first discovered in the Horsehead PDR \citep{Pety_IRAM30_2012}, and \citet{Guzman_Spatially_2015a} proposed that the high abundance of \lccchp\ ($\sim 10^{-11}$) in the Horsehead is due to the PDR chemistry which produces abundant $\rm C^+$ ions and possibly the photo-erosion of PAHs. 
\citet{Gerin_Molecular_2019} found that \lccchp\ is ubiquitous in diffuse gas, with an abundance of $\sim 7\times 10^{-11}$. 
The first detection of \lccchp\ in quiescent dense MC is reported by \citet{Cernicharo_Discovery_2022} who found $X({\rm C_3H^+})\sim 2.4\times 10^{-12}$ in the TMC-1 cloud. 
This value is lower than our observed value by an order of magnitude. 

\par

We note that the visual extinction of 3C391:NML is $A_{\rm V}\approx 6.2$ assuming $N_{\rm H}/A_{\rm V}=1.87\times 10^{21} \rm \  cm^{-2}\ mag^{-1} $ \citep{Bohlin_survey_1978}. 
In addition, we do not find any source of strong UV radiation close to the MC in the SIMBAD database\footnote{\url{https://simbad.u-strasbg.fr/simbad/}} \citep{Wenger_SIMBAD_2000}. 
Therefore, the chemistry of 3C391:NML is not likely to be dominated by UV photons \citep{Wolfire_Photodissociation_2022}. 
On the other hand, CRs, which can enhance the abundances of ionized species like $\rm C^+$, $\rm H_3^+$, $\rm CO^+$ and $\rm H_3O^+$, may provide advantages to explain the observed abundance and abundance ratios.

\subsubsection{Results of chemical simulation} \label{sec:model}

To further investigate the chemistry and constrain the cosmic-ray ionization rate in 3C391:NML, we present a chemical simulation using the \texttt{DNautilus 2.0} chemical code \citep{Taniguchi_Large-scale_2024a}. This code represents an updated version of \texttt{DNAUTILUS.1.0} as introduced in \citet{Majumdar_Chemistry_2017}. It is capable of simulating time-dependent abundances in two phases (treating the entire grain as homogeneous) and three phases (making a distinction between the surface and bulk of the grain).

\par

In \texttt{DNautilus 2.0}, deuteration is achieved up to the 14th largest atom-based molecule present in the kida.uva.2014 network, which is available in the KIDA database\footnote{\url{https://kida.astrochem-tools.org/}}. The deuteration routine used is described in \citet{Albertsson_New_2013}, resulting in 1606 gas species, 1472 grain-surface species, and 1472 grain-mantle species. These are connected by 83,715 gas-phase reactions, 10,967 reactions on grain surfaces, and 9,431 reactions in the grain mantles in \texttt{DNautilus 2.0}.

\par

The simulation consists of two steps. 
In step 1, we simulate the chemistry of a dense MC with low CR ionization rate per \hh\ ($\zeta= 10^{-17}\rm \ s^{-1}$) and low gas temperature $T=10\rm \ K$ for $t_1=0.1$ Myr and 1 Myr to mimic the chemical evolution of the MC before the supernova explosion. 
A grid of density $n_{\rm H}$ (10 values between $2\times 10^3$ and $2\times 10^6\rm \ cm^{-3}$) and C/O element ratio (0.6, 0.8, 1.0 and 1.2) is explored in step 1. 
Initially, all the abundances are in elemental form except for hydrogen and deuterium which appear in molecular form as H$_{2}$ and HD, respectively. Elements whose ionization potentials are less than 13.6 eV such as C, S, Si, Fe, Na, Mg, Cl and P appear in their first ionization states.
In step 2, we use the results of step 1 as input abundances, keep the density and the C/O ratio constant, and vary $\zeta$ ($10^{-17},\ 10^{-16},\ 10^{-15}$ and $10^{-14}\rm \ s^{-1}$) and $T$ (10 values between 10 and 50 K) to mimic the impact of CRs after the supernova explosion. 
We note that the CR ionization rate has non-negligible influence on the temperature of molecular gas \citep{Goldsmith_Molecular_1978}. 
However, it is hard to determine the exact values of gas temperature at different CR ionization rate analytically, and different calculations give different results \citep[e.g.,][]{Bayet_Chemistry_2011,Bisbas_Cosmic-ray_2017}. 
Therefore, we set these two parameters independent and fix the temperature throughout Step 2. Throughout the simulation, the cloud is assumed to be shielded from incident UV. 
But the secondary UV photons generated by CR radiation are considered, with $10^{4}$ secondary photons per CR field of 1.3$\times10^{-17}$ s$^{-1}$ \citep{Shen_Cosmic_2004a}. 
The physical parameters that are explored are listed in table \ref{tab:phy_parameters}. 
The initial elemental abundances for Step 1 are shown in Table \ref{tab:initial_abundances_step1} and the initial abundances of $\rm H_3^+$, $\rm H_2D^+$, \hcop, \dcop and CO for Step 2 for C/O ratio 0.8 in Table \ref{tab:initial_abundances_step2}.

\begin{deluxetable}{ccc}[!t]
\tablecaption{Physical parameters for step 1 and step 2\label{tab:phy_parameters}}
\tablehead{
\colhead{Parameter} & \colhead{Value used in step 1} & \colhead{Value used in step 2}
}
\startdata
$t_1$ (Myr) & 0.1, 1 & --- \\
$T$ (K) &  10 & 10--50\\
$n_{\rm {H}}$ (cm$^{-3}$) &  $2\times 10^3$--$2\times 10^6$ & $2\times 10^3$--$2\times 10^6$\\
$\log_{10}(\zeta/s^{-1})$ & $-17$ & $-17$, $-16$, $-15$, $-14$ \\
C/O ratio & 0.6, 0.8, 1.0, 1.2 & 0.6, 0.8, 1.0, 1.2
\enddata
\end{deluxetable}

\begin{deluxetable}{ll}[!t]
\tablecaption{Initial elemental abundances for step 1\label{tab:initial_abundances_step1}}
\tablehead{
\colhead{Element} & \colhead{Abundance relative to H} 
}
\startdata
H$_2$ &   $0.50$ \\
He    &   $9.00$($-2$) \\
N     &   $6.20$($-5$) \\
O     &   (2.83, 2.12, 1.70, 1.41)($-4$) \\
C$^+$ &   $1.70$($-4$) \\
S$^+$ &   $8.00$($-8$) \\
Si$^+$&   $8.00$($-8$) \\
Fe$^+$&   $3.00$($-9$) \\
Na$^+$&   $2.00$($-9$) \\
Mg$^+$&   $7.00$($-9$) \\
P$^+$ &   $2.00$($-10$) \\
Cl$^+$&   $1.00$($-9$) \\
F     &   $6.68$($-9$) \\
HD    &   $1.60$($-5$) 
\enddata
\end{deluxetable}

\begin{deluxetable*}{ccccccc}[ht]
\tablecaption{Relative abundance of some molecules w.r.t H for step 2 for C/O ratio = 0.8\label{tab:initial_abundances_step2}}
\tablehead{
\colhead{t$_{1}$ (Myr)} & \colhead{$n_{\rm H}$ (cm$^{-3}$)} & \colhead{$\rm H_3^+$} & \colhead{$\rm H_2D^+$} & \colhead{\hcop} & \colhead{\dcop} & \colhead{CO}
}
\startdata
\multirow{4}{*}{0.1} & 2 $\times$ 10$^{3}$ & 2.41($-9$) & 7.70($-11$) & 1.53($-9$) & 1.50($-11$) & 3.28($-5$) \\
& 2 $\times$ 10$^{4}$ & 2.59($-10$) & 1.17($-11$) & 3.55($-10$) & 2.67($-12$) & 4.62($-5$) \\
& 2 $\times$ 10$^{5}$ & 7.58($-10$) & 8.55($-10$) & 8.96($-11$) & 1.26($-10$) & 4.11($-7$) \\
& 2 $\times$ 10$^{6}$ & 2.14($-10$) & 2.89($-10$) & 1.85($-12$) & 2.94($-12$) & 7.90($-9$) \\
\midrule
\multirow{4}{*}{1.0} & 2 $\times$ 10$^{3}$ & 4.90($-9$) & 3.02($-10$) & 7.55($-9$) & 2.10($-10$) & 7.34($-5$) \\
& 2 $\times$ 10$^{4}$ & 7.20($-9$) & 2.88($-9$) & 4.72($-10$) & 1.34($-10$) & 1.27($-6$) \\
& 2 $\times$ 10$^{5}$ & 4.07($-9$) & 1.14($-9$) & 3.76($-11$) & 5.29($-12$) & 7.50($-8$) \\
& 2 $\times$ 10$^{6}$ & 1.36($-9$) & 3.39($-10$) & 3.98($-12$) & 3.76($-13$) & 7.57($-9$) \\
\enddata
\end{deluxetable*}

\begin{figure*}[ht]
\centering
\includegraphics[width=0.99\textwidth]{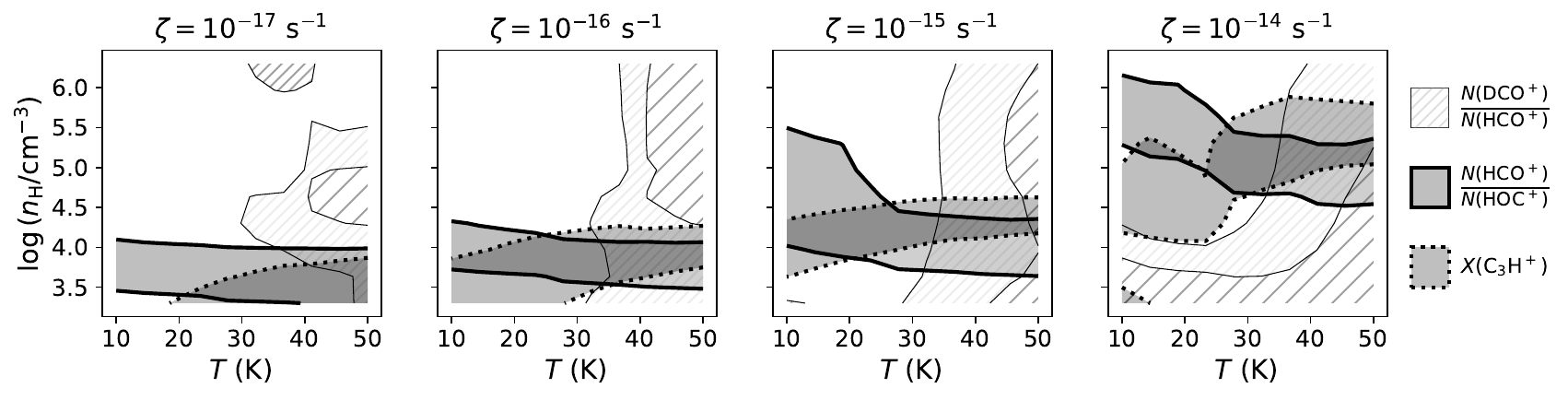}
\includegraphics[width=0.99\textwidth]{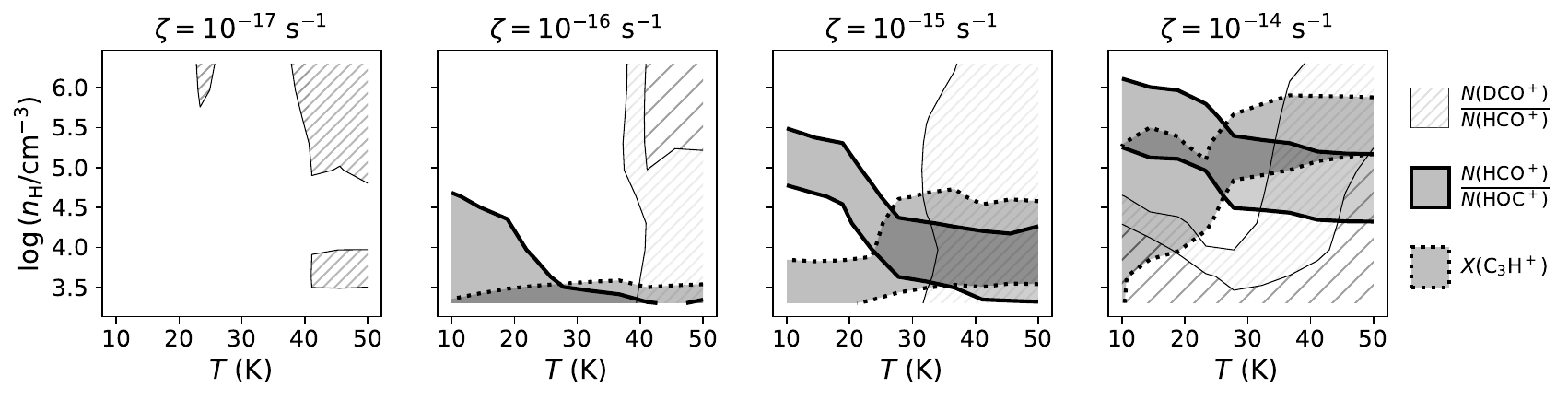}
\caption{
Results of the chemical simulation for \dcoponhcop, \hcoponhocp\ and \xlccchp\ with $t_1=0.1$ (upper panel) and 1 (Lower panel) Myr and $\rm C/O=0.8$.
Each column shows the results with fixed $\zeta$ (CR ionization rate per \hh) from $10^{-17}$ to $10^{-14} \rm \ s^{-1}$. 
The hatched and shaded regions shows the combinations of parameters which can reproduce the observational results within a factor of 2. 
As denoted in the labels in the right, the densely hatched regions with solid thin bounds show \dcoponhcop\ (while the sparsely hatched regions mean that the modeled \dcoponhcop\ is lower than half of the estimated upper limits obtained from observation), the shaded regions with solid thick bounds show \hcoponhocp, and the shaded regions with dotted bounds show \xlccchp. 
\label{fig:model}}
\end{figure*}

\begin{table*}[ht]
\centering
\caption{Major Chemical Reactions for HCO$^+$, HOC$^+$, DCO$^+$, and CO}
\label{tab:major_chemical_reactions}
\begin{tabular}{|c|c|c|}
\hline
\textbf{Species} & \textbf{Production Reactions} & \textbf{Destruction Reactions} \\
\hline
HCO$^+$ &
CO + H$_3^+$ $\rightarrow$ H$_2$ + HCO$^+$ & HCO$^+$ + e$^-$ $\rightarrow$ H + CO \\
& & HCN + HCO$^+$ $\rightarrow$ CO + HCNH$^+$ ($\zeta \lesssim 10^{-16}$ s$^{-1}$ and $T\gtrsim$ 43 K)\\
\hline
HOC$^+$ &
CO + H$_3^+$ $\rightarrow$ H$_2$ + HOC$^+$ & HOC$^+$ + H$_2$ $\rightarrow$ HCO$^+$ + H$_2$ \\
& H$_2$O + C$^+$ $\rightarrow$ H + HOC$^+$ ($n_{\rm H} \lesssim$ 2$\times 10^{4}$ cm$^{-3}$) & HOC$^+$ + e$^-$ $\rightarrow$ H + CO ($n_{\rm H} \approx$ 2 $\times$ 10$^{3}$ cm$^{-3}$) \\
& H$_2$ + CO$^+$ $\rightarrow$ H + HOC$^+$  & \\
\hline
DCO$^+$ &
HCO$^+$ + D $\rightarrow$ DCO$^+$ + H & DCO$^+$ + HCN $\rightarrow$ HDCN$^+$ + CO (T $\gtrsim$ 43 K)\\
&CO + CH$_4$D$^+$ $\rightarrow$ CH$_4$ + DCO$^+$ & DCO$^+$ + e$^-$ $\rightarrow$ D + CO \\
\hline
CO &
HCO$^+$ + e$^-$ $\rightarrow$ CO + H & CO + H$_3^+$ $\rightarrow$ H$_2$  + HCO$^+$ \\
&CO$_{\rm ice}$ $\rightarrow$ CO (T $\gtrsim$ 23 K) & CO $\rightarrow$ CO$_{\rm ice}$ (T $\gtrsim$ 23K) \\
& & CO + He$^+$ $\rightarrow$ He + O + C$^+$ \\
\hline
\end{tabular}
\end{table*}

The evolution time for step 2 is chosen according to the age of 3C391. 
With various scenarios such as Sedov evolution, cloud evaporation, and radiative phase, the age of SNR 3C391 was estimated within a range between 4-19 kyr \citep{Reynolds_High-Resolution_1993,Rho_ROSAT_1996,Chen_ASCA_2001,Chen_Chandra_2004,Leahy_Evolutionary_2018}. 
We extract the abundances of the species in step 2 at 8000 yr, which is an intermittent value. 
The results of the simulation are shown in Figure \ref{fig:model} for both $t_1=0.1$ and 1 Myr. 
We note that the difference in results with different C/O ratio is negligible. 
So we only show the results with $\rm C/O=0.8$ as a representative in Figure \ref{fig:model}. 
The difference between the results with $t_1=0.1$ and 1 Myr is due to the different level of depletion in Step 1, and thus the different initial condition in Step 2. 
Although the \hcoponco\ abundance ratio can be used to estimate the CR ionization rate (Equation \ref{eq:CRIR}), here we do not use it for our simulation because CO traces less dense gas compared with other species (See Section \ref{sec:nonLTE} and \citet{Shirley_Critical_2015}).

\par

The simulation fails to reproduce the observed values of \hcsponcs\ (0.14--0.18), with typical simulated values ${\lesssim 10^{-2}}$ in any combination of parameters. 
We attempted to reproduce the observed N(HCS+)/N(CS) ratios by assuming different initial sulfur abundances. However, all models do not match the observed values. 
This may result from some chemical effects of sulfur, for example, the form of sulfur depletion in dark MCs \citep{Vidal_reservoir_2017}, which are not considered in our chemical model. 
Due to this sulfur conundrum, we do not show the results here. 

\par

The two abundance ratios are both related to the \hcop\ molecule. 
The formation of \hcop\ is mainly through:
\begin{equation}
\label{eq:hco+p}
    \rm CO + H_3^+  \rm \longrightarrow HCO^+ + H_2 \\
\end{equation}
in most of the physical conditions whereas the formation reactions of \hocp\ including reactions (\ref{react:c++h2o}) and (\ref{react:co++h2}) depend on the specific value of $\zeta$, $n_{\rm H}$ and $T$. 
The major production and destruction pathways of \hcop\ and \hocp\ are listed in Table \ref{tab:major_chemical_reactions}.
We note that the abundance ratio \hcoponhocp\ exhibits a rather regular dependency on different physical parameters at $T\lesssim 40\rm \ K$---decreased with higher $\zeta$ and lower $n_{\rm H}$, while decreased with higher $T$.
We therefore propose that the low \hcoponhocp\ abundance ratio may be a good tracer of enhanced CR ionization rate in dark unshocked cloud associated with SNRs if other physical parameters are well constrained by observation. 

\par

The molecule \lccchp, as a carbon chain species, is expected to be sensitive to the C/O ratio \citep[e.g.,][]{Pratap_Study_1997}. 
However, we find that the dependence of \xlccchp\ on the C/O ratio is not very significant, which is also reported by \citet{Loison_gas-phase_2014}, and does not affect our conclusions. 
We note that the \xlccchp\ is higher when the CR ionization rate is higher. 

\par

In general, we find that four combinations of parameters can reproduce the three observed values: 
(1) $t_1=0.1\rm \ Myr$, $n_{\rm H}\sim 10^{3.5} \ \rm cm^{-3}$, $T\sim50\rm \ K$, $\zeta\sim 10^{-17}\rm \ s^{-1}$; 
(2) $t_1=0.1\rm \ Myr$, $n_{\rm H}\sim 10^{4} \ \rm cm^{-3}$, $T\sim35\rm \ K$, $\zeta\sim 10^{-16}\rm \ s^{-1}$; 
(3) $n_{\rm H}\sim 10^{4} \ \rm cm^{-3}$, $T\sim35\rm \ K$, $\zeta\sim 10^{-15}\rm \ s^{-1}$; and
(4) $n_{\rm H}\sim 10^5\ \rm cm^{-3}$, $T\sim 40\rm \ K$, $\zeta\sim 10^{-14}\rm \ s^{-1}$. 
Combinations 1 and 2 are only valid for $t_1=0.1\rm \ Myr$, while 3 and 4 are valid for both values of $t_1$. 
The densities in all the four combinations are within the range obtained from our non-LTE analysis of CO, \occthreehtwo, and CS (see Section \ref{sec:nonLTE}). 

\par

However, as we have mentioned above, the gas temperature is dependent on the CR ionization rate because of the heating effect of CRs. 
The calculation of \citet{Bayet_Chemistry_2011} shows that the gas temperatures are $\sim 10$, $\sim 20$, $\sim 40$ and $\sim 80$ K for $\zeta\sim 10^{-17},\ 10^{-16},\ 10^{-15}$ and $10^{-14} \rm \ s^{-1}$, respectively, while \citet{Bisbas_Cosmic-ray_2017} predicts $\sim 10$, $\sim 10$, $\sim 20$ and $\sim 40$ K correspondingly. 
Comparing these values with our simulation results, we find that combinations 1 and 2 are not compatible with the CR heating as they require higher temperatures than the values predicted by \citet{Bayet_Chemistry_2011} and \citet{Bisbas_Cosmic-ray_2017}, while the slightly elevated temperatures in combinations 3 and 4 can be naturally explained by CR heating. 

\par

We note that SNR 3C391 is a bright source in X-rays which can penetrate deeper into the MCs compared with UV photons \citep{Wolfire_Photodissociation_2022} and can heat the molecular gas deep inside the MCs. 
To evaluate whether X-rays can provide the required heating for combinations 1 and 2, we here estimate and compare the heating rates of X-rays and CRs. 
For the X-ray emission with a photon index $\alpha$, the X-ray heating rate per unit volume is \citep{Maloney_X-Ray_1996a}: 
\begin{equation}
    \Gamma_{\rm X}\approx 3\times 10^4 \left(\frac{f_{\rm h}}{0.3} \right) \left(\frac{n_{\rm H}}{10^5\rm \ cm^{-3}} \right) H_{\rm X} 
\end{equation}
where $f_{\rm h}\approx 0.3$ is the fraction of primary photoelectron energy that goes into gas heating, and 
\begin{equation} \label{eq:HX}
    H_{\rm X}\approx \frac{3\sigma_0C_\alpha F_{\rm X}}{8\tau_1^{\phi+1}}S(\tau_1),
\end{equation}
where $\sigma_0 = 2.6\times 10^{-22}\rm \ cm^2$, 
$C_\alpha = (1-\alpha)/(E_{k,\rm max}^{1-\alpha} - E_{k,\rm min}^{1-\alpha})$ for $\alpha\neq1$, 
$E_{k,\rm min}$ and $E_{k,\rm max}$ are the minimum and maximum energies in units of keV, 
$F_{\rm X}$ is the X-ray flux at the target point, 
$\phi=3(\alpha-1)/8$, 
the optical depth at 1 keV from the X-ray source to 3C391:NML is $\tau_1=2.6N_{22}$, 
and $S(\tau_1)\approx 1$ for $10^{-2}\lesssim \tau_1 \lesssim 10^4$. 
For 3C391, the X-ray luminosity is $2.3\times 10^{36}\rm \ erg \ s^{-1}$ in 0.5--10 keV, and the mean distance between 3C391 and 3C391:NML is 7 pc \citep{Chen_Chandra_2004}. 
The attenuating column density $N_{\rm H}$ is assumed to be $10^{22} \rm \ cm^{-2}$ so that $N_{22}=1$. 
We note that $\Gamma_{\rm X}$ depends on the X-ray photon index $\alpha$, but the X-ray spectrum of 3C391 is dominated by a thermal component which cannot be fitted simply with a power law. 
Assuming $\alpha\approx3.3$ which is adopted by \citet{Zhou_Molecular_2018} in SNR Cas A, and $n_{\rm H}=10^4\rm \ cm^{-3}$, we get $\Gamma_{\rm X}\sim 4\times 10^{-27}\rm \ erg\ cm^{-3}\ s^{-1}$. 
Changing the value of $\alpha$ does not affect the basic conclusion that $\Gamma_{\rm X}<10^{-26} \rm \ erg\ cm^{-3}\ s^{-1}$. 

\par

For the heating rate of CRs, we adopt the equations in \citet{Goldsmith_Molecular_1978} and the assumptions in \citet{Zhou_Molecular_2018}: 
\begin{equation}
\begin{aligned}
    \Gamma_{\rm CR} \approx\ &3.2\times 10^{-25}\left(\frac{n_{\rm H_2}}{10^3\rm \ cm^{-3}}\right)\times  \\
    &\left(\frac{\zeta}{10^{-17}\rm \ s^{-1}}\right)\ \rm erg\ cm^{-3}\ s^{-1}. 
\end{aligned}
\end{equation} 
For combinations 1 and 2 in our simulation, we get $\Gamma_{\rm CR}\approx 10^{-24}$ and $2\times 10^{-23}\rm\ erg\ cm^{-3}\ s^{-1}$, respectively, which are significantly higher than the X-ray heating rate. 
Considering that there is no evidence of other heating sources such as star formation activities \citep{Urquhart_ATLASGAL_2018}, shocks (introduced in Section \ref{sec:intro}) and external UV radiation (discussed in Section \ref{sec:abun}), we suggest that combinations 1 and 2 cannot be used to reproduce the observation. 

\par

For combinations 3 and 4, gas temperatures of $\sim 35$ and $\sim 40$ K are required. 
Due to the large scatter in the relation between $T$ and the line ratio $I({\rm H^{13}CN})/I({\rm HN^{13}C})$ \citep{Pazukhin_H13CN-HN13C_2022} and the uncertainty in the non-LTE analysis in Section \ref{sec:nonLTE} (for example, the unknown spatial distribution of the molecular gas inside the telescope beam), the moderately enhanced temperatures in combinations 3 and 4 are indeed possible. 
Adopting a temperature of 40 K does not result in a significant change in the molecular column densities (See Section \ref{sec:nonLTE}). 
In conclusion, enhanced CR ionization rate ($\sim 10^{-15}$ or $10^{-14}\rm \ s^{-1}$) is required to reproduce the observation according to the simulation. 

\subsection{Enhanced CR ionization rate associated with SNR 3C391}

In the previous section, we find an enhanced CR ionization rate associated with SNR 3C391.
However, X-ray can also induce chemistry similar to CRs \citep{Viti_Molecular_2017}. 
\citet{Chen_Chandra_2004} found that the X-ray ionization rate of the position towards an 1720 MHz OH maser which is close to 3C391:NML (See Figure \ref{fig:1}) is $\zeta_{\rm X}\sim 2\times 10^{-15} \rm \ s^{-1}$, but their estimation may overestimate the X-ray ionization rate because the equation assumes the X-ray spectrum comprises only photons of 1 keV \citep{Wardle_Enhanced_1999}, which have the strongest ionization effect.  
For a more realistic consideration, we adopt the equations in \citet{Maloney_X-Ray_1996a}:
\begin{equation}
    \zeta_{\rm X}\approx 1.83\times 10^{10} \frac{f_{i}}{0.4}H_{\rm X},
\end{equation}
where $f_i\approx 0.4$ is the fraction of primary photoelectron energy that goes into ionization and $H_{\rm X}$ has been defined in Equation \ref{eq:HX}. 
Based on the assumptions in the previous section, we get $\zeta_{\rm X}\sim 3\times 10^{-20}\rm \ s^{-1}$ which is significantly lower than the CR ionization rate. 
Changing the value of $\alpha$ does not affect the basic conclusion that $\zeta_{\rm X}<10^{-19} \rm \ s^{-1}$. 
Although this is just a rough estimation on the order of magnitude of $\zeta_{\rm X}$, we conclude that the chemistry in 3C391:NML is dominated by CRs instead of X-ray photons. 

\par

The obtained CR ionization rate, $\zeta\sim 10^{-15}$ or $10^{-14}\rm \ s^{-1}$, is $\sim 2$--3 orders of magnitude higher than typical values in MCs \citep{Glassgold_Model_1974a,Caselli_Ionization_1998}. 
The CR ionization rate of $\sim 10^{-15}\rm s^{-1}$ is similar to the values found in MCs associated with SNRs W51C \citep{Ceccarelli_Supernova-enhanced_2011} and W28 \citep{Vaupre_Cosmic_2014}. 
If $\zeta\sim 10^{-14}\rm \ s^{-1}$ is true, this is even higher by an order of magnitude than the values in MCs associated with other SNRs. 
This extremely high value of CR ionization rate has been found in the CMZ of our Galaxy \citep[e.g.,][]{Oka_Central_2019}, the solar-type protostar OMC-2 FIR4 \citep{Ceccarelli_Herschel_2014,Fontani_Seeds_2017}, and CMZ of other galaxies like NGC 253 \citep{Holdship_Energizing_2022}. 
The origin of the extremely high CR ionization rate in the CMZ of our Galaxy arises from a combination of energetic activities including SNRs and colliding winds of massive stars \citep[e.g.,][]{Yusef-Zadeh_Origin_2007}, whereas \citet{Lattanzi_SOLIS_2023} proposed that the high CR ionization rate in OMC-2 FIR4 is due to the CRs accelerated in the jet shock of the young protostar. 
However, no star formation activity is found to be associated with 3C391:NML \citep{Urquhart_ATLASGAL_2018}. 
Therefore, it is more likely that the extremely high $\zeta$ obtained in 3C391:NML is originated from the SNR. 
Extrapolating the high energy CR spectrum inferred from the $\gamma$-ray observation to low energies, \citet{Schuppan_Cosmic-ray-induced_2012} found that SNR 3C391 can induce $\zeta\sim 10^{-14}\rm \ s^{-1}$ in its surrounding MCs. 
According to their calculation, a possible reason why 3C391 can induce higher CR ionization rate than other SNRs is that the deduced proton flux of 3C391 at $<1\rm \ GeV$ is higher than those of other SNRs, which means the MCs adjacent to 3C391 is exposed to more ionizing low-energy CR protons. 
Yet this extrapolation is not exempt from problems, because the CRs leading to ionization have too low energy \citep[$\lesssim 280\rm \ MeV$,][]{Padovani_Cosmic-ray_2009} to produce detectable $\gamma$-ray emission by available instruments.

\section{Conclusion} \label{sec:con}

In this paper, we performed a W band (71.5--90 GHz) line survey with the Yebes 40 m radio telescope towards an unshocked molecular cloud of SNR 3C391, which we call 3C391:NML. 
Our main conclusions are summarized as follows: 

\par

1. We detected 18 molecules in the line survey. 
The line profile of the \hcop\ $J=1$--0 line exhibits three components with different linewidths, while most molecular lines only exhibit a narrow component and some exhibit two components. 

\par

2. Assuming $T=20\rm \ K$, we estimated the physical parameters of 3C391:NML using the SpectralRadex code with the CO, \occthreehtwo\ and CS lines. 
The density is estimated to be $n_{\rm H_2}\sim 1.4\times 10^3 \rm \ cm^{-3}$ from the CO lines, and $\sim 2\times 10^5\rm \ cm^{-3}$ from the \occthreehtwo\ and CS lines. 
The estimated $N_{\rm H_2}$ of the MC is $\sim 5.8\times 10^{21}\rm \ cm^{-2}$. 

\par

3. Using the analytic equations reported by \citet{Vaupre_Cosmic_2014}, we estimated the CR ionization rate of 3C391:NML is $\gtrsim 2.7\times 10^{-14}\rm \ s^{-1}$ with the abundance ratio \hcoponco\ and an upper limit of \dcoponhcop. 
However, we caution on adopting this value because chemical equilibrium, which is a prerequisite of using the equations, is not necessarily reached in the MC considering the 4--19 kyr age of 3C391. 

\par

4. We found some unusual abundance and abundance ratios compared with typical values in quiescent dense MCs. 
They are: $N({\rm HCO^+})/N({\rm HOC^+})\sim 160\text{--}180$ lower than typical values, $N({\rm HCS^+})/N({\rm CS})\sim 0.14\text{--}0.18$ higher than typical values, and $X(l\text{-}\rm C_3H^+ )\sim3.1\text{--}6.0\times 10^{-11}$ higher than typical values by an order of magnitude. 
These can be attributed to the chemistry induced by CRs. 

\par

5. Using the \texttt{DNautilus 2.0} chemical model equipped with deuterium chemistry, we present a chemical simulation to explain the observed abundance and abundance ratios. 
We found that two combinations of parameters: $n_{\rm H}\sim 10^{4} \ \rm cm^{-3}$, $T\sim35\rm \ K$, $\zeta\sim 10^{-15}\rm \ s^{-1}$ and $n_{\rm H}\sim 10^5\ \rm cm^{-3}$, $T\sim 40\rm \ K$, $\zeta\sim 10^{-14}\rm \ s^{-1}$ can reproduce the observation. 
The moderately elevated temperatures can be naturally explained by CR heating. 
The obtained CR ionization rate is higher than typical values in MCs by 2--3 orders of magnitude. 
If $\zeta\sim 10^{-14}\rm \ s^{-1}$ is true, this is higher than those in MCs associated with other SNRs by an order of magnitude, which may be because 3C391 is exposed to more ionizing low-energy CR protons compared with other SNRs.

\begin{acknowledgments}
The authors thank the anonymous referee for helpful comments and suggestions. 
T.-Y. T. thanks Siyi Feng, Thomas Bisbas, Yichen Sun and Xiao Zhang for helpful discussions. 
L.M. acknowledges the financial support of DAE and DST-SERB research grants (SRG/2021/002116 and MTR/2021/000864) of the Government of India.
Y.C. acknowledges the support from NSFC grants Nos. 12173018, 12121003 and 12393852. 
P.Z. acknowledges the support from NSFC grant No. 12273010. 
M S. G. has carried out the observations and the first inspection of the data quality. 

\par

This article is based on observations carried out with the Yebes 40 m telescope (project code: 23A021). The 40 m radio telescope at Yebes Observatory is operated by the Spanish Geographic Institute (IGN; Ministerio de Transportes, Movilidad y Agenda Urbana). 
This research has also made use of the SIMBAD database, operated at CDS, Strasbourg, France.

\end{acknowledgments}

\vspace{5mm}

\facilities{Yebes, No: 45m, JCMT}

\software{astropy \citep{AstropyCollaboration_Astropy_2018, AstropyCollaboration_Astropy_2022},  
          Spectral-cube \citep{Ginsburg_Radio_2015}, 
          GILDAS (Gildas Team, \url{https://www.iram.fr/IRAMFR/GILDAS/}), 
          Montage (\url{http://montage.ipac.caltech.edu/}, 
          Matplotlib (\url{https://matplotlib.org}))
          }

\appendix

\section{Full spectrum}
\begin{figure*}[ht]
\centering
\includegraphics[width=\textwidth]{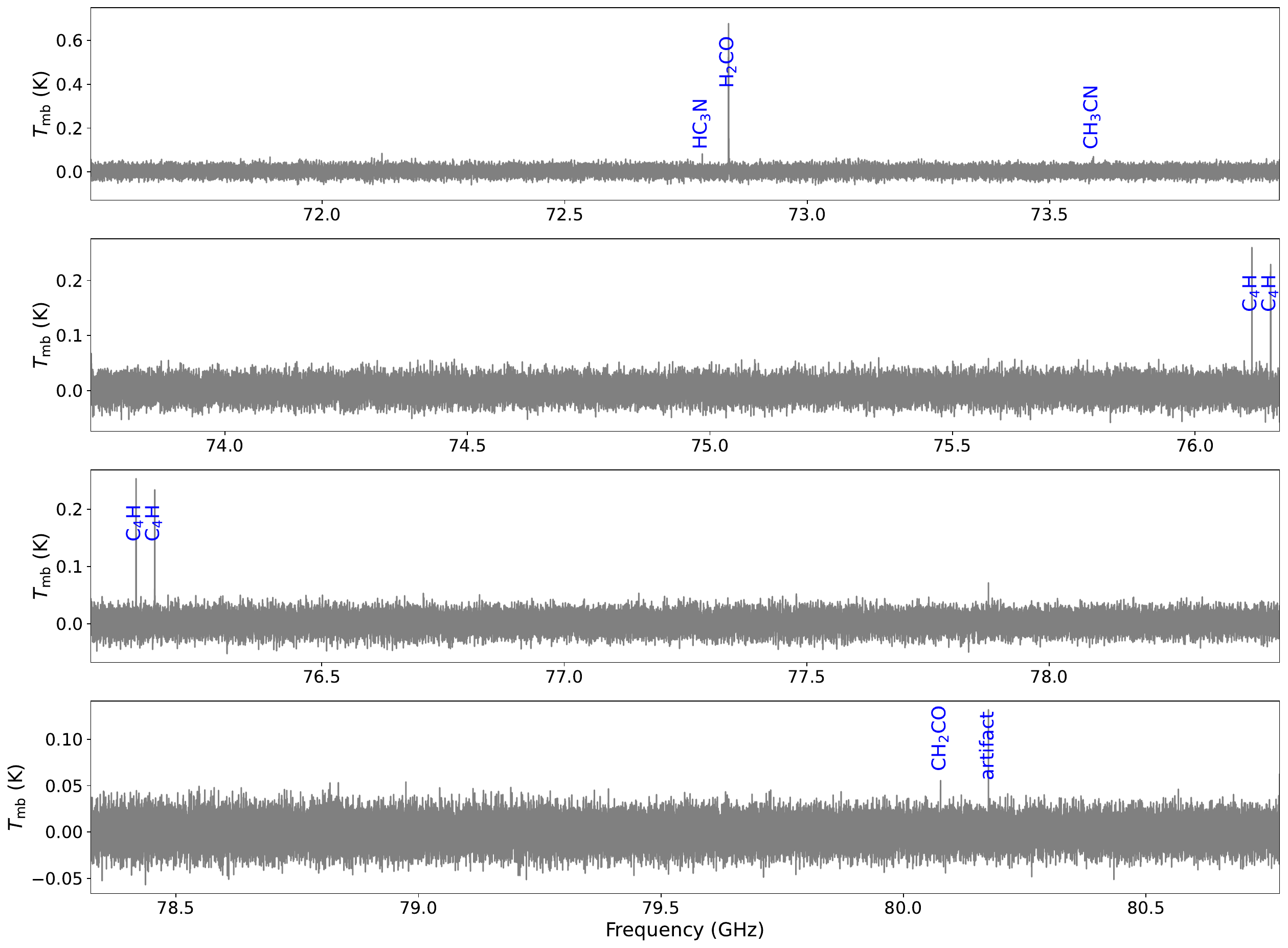}
\caption{ 
Full spectrum in the entire frequency range covered by eight spectrometers. Every two adjacent spectrometers have overlaps in frequency. The detected species and strong artifacts are marked in blue. 
\label{fig:full}}
\end{figure*}

\begin{figure*}[ht]
\centering
\figurenum{7}
\includegraphics[width=\textwidth]{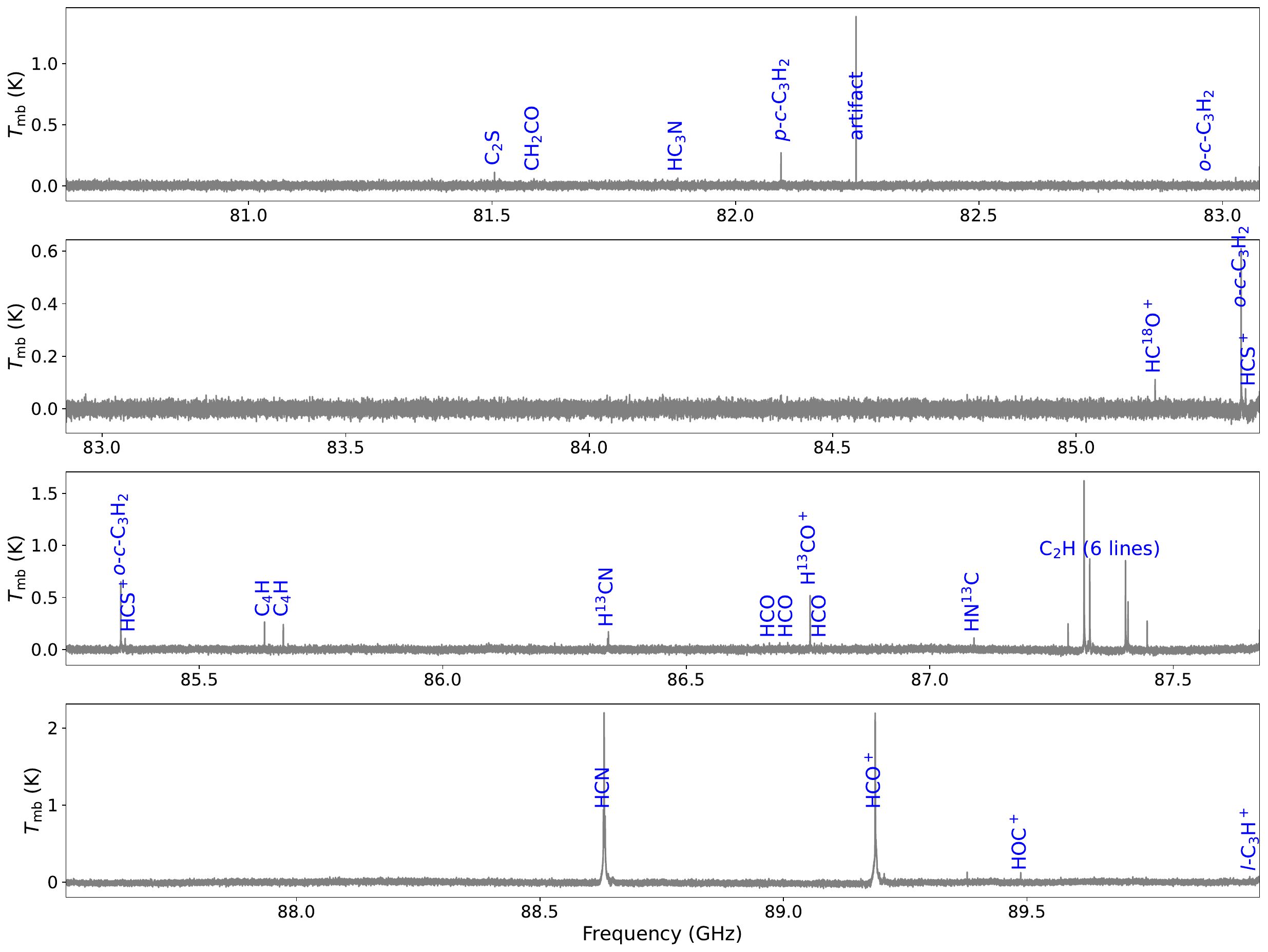}
\caption{ \textit{Continued.}
\label{fig:full2}}
\end{figure*}

\bibliography{3C391_3mm_survey_article}{}
\bibliographystyle{aasjournal}

\end{CJK*}
\end{document}